\documentclass[aps,pra,twocolumn,showpacs]{revtex4}
\usepackage{graphicx} 
\usepackage{amsmath}
\usepackage{bm}
\usepackage{dcolumn}
\begin{document}

\title{Excitation energies, polarizabilities, multipole transition rates,
and  lifetimes   of ions along the francium isoelectronic sequence
}

\author{ U. I. Safronova }
\email{usafrono@nd.edu}\altaffiliation{ On  leave  from ISAN,
Troitsk, Russia} \affiliation{Physics Department, University of
Nevada, Reno, NV
 89557}

\author{W. R. Johnson}
\email{johnson@nd.edu} \homepage{www.nd.edu/~johnson}
\affiliation{Department of Physics,  University of Notre Dame,
Notre Dame, IN 46556}

\author{M. S. Safronova}
\email{msafrono@udel.edu} \affiliation {Department of Physics and
Astronomy, 223 Sharp Lab,
 University of Delaware, Newark, Delaware 19716}

\date{\today}
\begin{abstract}
Relativistic many-body perturbation theory is applied to study
properties of ions of the francium isoelectronic sequence.
Specifically, energies of the $7s$, $7p$, $6d$, and $5f$ states of
Fr-like ions with nuclear charges $Z = 87 - 100$ are calculated
through third order; reduced matrix elements, oscillator
strengths, transition rates, and lifetimes are determined for
$7s-7p$, $7p -6d$, and $6d -5f$ electric-dipole transitions; and
$7s - 6d$, $7s - 5f$, and $5f_{5/2}\ - 5f_{7/2}$ multipole matrix
elements are evaluated to obtain the lifetimes of low-lying
excited states. Moreover, for the ions $Z = 87 - 92$ calculations
are also carried out using the relativistic all-order
single-double method, in which single and double excitations of
Dirac-Fock wave functions are included to all orders in
perturbation theory.  With the aid of the SD wave functions, we
obtain accurate values of energies, transition rates, oscillator
strengths, and the lifetimes of these six ions.  Ground state
scalar polarizabilities in Fr~I, Ra~II,  Ac~III, and Th~IV are
calculated using relativistic third-order and all-order methods.
Ground state scalar polarizabilities for other Fr-like ions are
calculated using a relativistic second-order method. These
calculations provide a theoretical benchmark for comparison with
experiment and theory.
 \pacs{31.15.Ar, 31.15.Md, 32.10.Fn, 32.70.Cs}
\end{abstract}
\maketitle

\section{Introduction}
\begin{figure}[tbp]
\centerline{\includegraphics[scale=0.4]{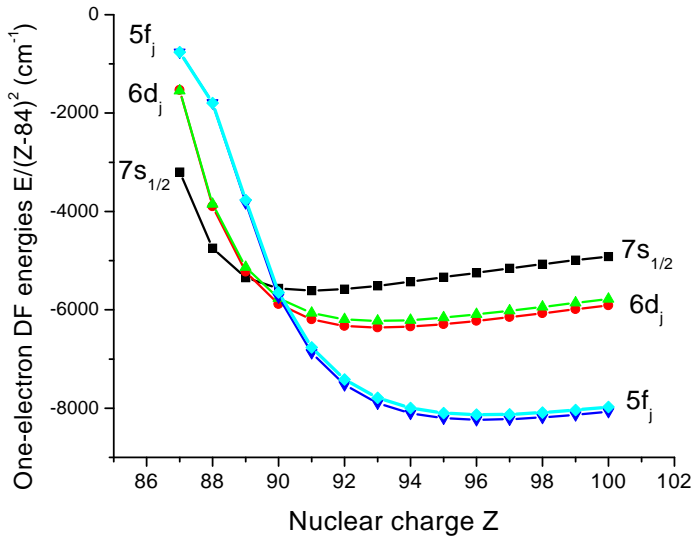}}
\caption{Dirac-Fock one-electron energies ($E/(Z-84)^2$ in
cm$^{-1}$) for the $5f_j$, $6d_j$, and $7s_{1/2}$  states of
Fr-like ions as functions of $Z$.}
 \label{fig-e0}
\end{figure}

A detailed investigation of radiative parameters for electric
dipole (E1) transitions in Fr-like ions with $Z$ = 89--92 was
presented recently by \citet{osc-ra}.  The electronic structure of
Fr-like   ions consists of a single $nl$ electron outside of a
core with completely filled   $n$=1, 2, 3, 4 shells and $5s$, $5p$,
$5d$, $6s$, and $6p$ subshells. In Fig.~\ref{fig-e0}, we plot
one-electron DF energies of {\it valence} $5f$,   $6d$, and
$7s$ states as functions of $Z$. We find that the valence $7s$
orbital is more tightly bound than the $5f$ and $6d$ orbitals
at low stages of ionization ($Z$ = 87--89), while the  $5f$
and $6d$ orbitals are more tightly bound for highly ionized
cases ($Z \geq$~90). Competition between the $5f$, $6d$,
and $7s$ orbitals leads to problems for calculations, making
it difficult to obtain very accurate excitation energies and line
strengths for the transitions between the low-lying $5f$,
$6d$, and $7s$ states.

Relativistic Hartree-Fock and Dirac-Fock atomic structure codes
were used in Ref.~\cite{osc-ra} to perform calculations of
radiative transition rates and oscillator strengths for a limited
number of transitions using the energies given by \citet{expt},
where experimental values were given for the $7p$, $8p$,
$nd$ (with $n$ =10--33), and $ns$ (with $n$ =12--31)
levels of neutral Fr, for 24 levels for Fr-like Th, and for seven
levels of Fr-like Ac and U. Adopted energy level values in the
$ns$, $np$, and $nd$ ($n\leq$ 30) series of neutral
francium were presented by  \citet{biemont-fr}. Experimental
measurements of energy levels of the $8s$ and $7d$
states in Fr~I were reported recently in Refs.~\cite{fr-8s,fr-7d}.
The experimental energies of 13 levels of Fr-like Ra were reported
in the NIST compilation \cite{web-nist}.

Lifetime measurements for neutral francium were presented in
Refs.~\cite{fr-97,fr-98,fr-00,fr-03,fr-05} for the $7p$, $6d$,
$9s$, and $8s$ levels. In those papers, experimental
measurements were compared with {\it ab-initio} calculations
performed by \citet{dip3},
 \citet{dzuba-95,dzuba-01}, \citet{safr-alk}, and \citet{safr-fr}.
 Third-order many-body perturbation
theory was used in Ref.~\cite{dip3} to obtain the E1
transition amplitude for neutral alkali-metal atoms. The
correlation potential method and the Feynman diagram technique were
used in Refs.~\cite{dzuba-95,dzuba-01} to calculate E1 matrix
elements in neutral francium and Fr-like radium. Atomic properties
of Th~IV ion were studied by ~\citet{thiv} using a relativistic
all-order method.

The present third-order calculations of excitation energies of the
$7s$, $7p$, $6d$, and $5f$ states in Fr-like ions with nuclear
charges $Z = 87 - 100$ start from a closed-shell Dirac-Fock
potential for the 86 electron radon-like core. We note that Th~IV
is the first ion in the francium isoelectronic sequence with a [Rn]$5f_{5/2}$
ground state instead of the [Rn]$7s_{1/2}$ ground state as for
Fr~I, Ra~II, and Ac~III. Correlation corrections become very large
for such systems as was demonstrated by \citet{igor}, where the
ratio of the second-order and DF removal energies for the
[Xe]$4f_{5/2}$ ground state in Ce~IV and Pr~V were
 18\% and 11\%, respectively.

In the present paper, dipole matrix elements are calculated using both
relativistic many-body perturbation theory, complete through
third-order, and the relativistic all-order method restricted to
single and double (SD) excitations. Such calculations permit one
to investigate the convergence of perturbation theory and estimate
the theoretical error in predicted data. To obtain lifetime
predictions, multipole matrix elements for $7s - 6d$, $7s - 5f$
and $5f_{5/2}\ - 5f_{7/2}$ transitions are also evaluated.
Additionally, scalar polarizabilities for the $7s_{1/2}$ ground
state in Fr~I, Ra~II, and Ac~III are calculated using relativistic
third-order and SD methods. Finally, scalar polarizabilities of the
$5f_{5/2}$ ground state of Fr-like ions with nuclear charge $Z$ =
91--100 are calculated in second-order MBPT.

\begin{table*}
\caption{\label{tab1} Zeroth-order (DF), second-, and third-order
Coulomb correlation energies $E^{(n)}$,  single-double Coulomb
energies $E^\text{{SD}}$, $E^{(3)}_\text{{extra}}$,
 first-order Breit and second-order Coulomb-Breit
 corrections $B^{(n)}$  to the energies of  Fr-like systems.
The total energies are $E^{(3)}_\text{ tot} = E^{(0)} + E^{(2)}
+E^{(3)}+ B^{(1)} + B^{(2)}$ + $E_{LS}$, $E^\text{SD}_\text{tot} =
E^{(0)} + E^\text{{SD}} + E^{(3)}_\text{{extra}} + B^{(1)} +
B^{(2)}$ + $E_{LS}$. Units: cm$^{-1}$.}
\begin{ruledtabular}\begin{tabular}{lrrrrrrrrrr}
\multicolumn{1}{c}{$nlj$ } & \multicolumn{1}{c}{$E^{(0)}$} &
\multicolumn{1}{c}{$E^{(2)}$} & \multicolumn{1}{c}{$E^{(3)}$} &
\multicolumn{1}{c}{$B^{(1)}$} & \multicolumn{1}{c}{$B^{(2)}$} &
\multicolumn{1}{c}{$E_{LS}$} & \multicolumn{1}{c}{$E^{(3)}_\text{
tot}$} & \multicolumn{1}{c}{$E^\text{{SD}}$} &
\multicolumn{1}{c}{$E^{(3)}_\text{{extra}}$} &
\multicolumn{1}{c}{$E^\text{{SD}}_\text{ tot}$} \\
\hline
\multicolumn{11}{c}{Fr~I, $Z$ = 87}\\
$7s_{1/2}$&   -28767&  -4763&  1737&   67& -131&  13&  -31845&    -4658&  724&  -32753 \\
$7p_{1/2}$&   -18855&  -1847&   546&   30&  -35&   0&  -20162&    -1991&  255&  -20597 \\
$7p_{3/2}$&   -17655&  -1346&   391&   19&  -30&   0&  -18621&    -1433&  184&  -18915 \\
$6d_{3/2}$&   -13807&  -2424&   711&   19&  -62&   0&  -15563&    -3180&  326&  -16990 \\
$6d_{5/2}$&   -13924&  -2240&   616&   15&  -61&   0&  -15594&    -2854&  288&  -16744 \\
$8s_{1/2}$&   -12282&  -1053&   396&   17&  -32&   2&  -12952&     -925&  162&  -13058 \\
$8p_{1/2}$&    -9240&   -554&   171&   10&  -13&   0&   -9625&     -551&   77&   -9716 \\
$8p_{3/2}$&    -8811&   -421&   128&    7&  -11&   0&   -9108&     -421&   58&   -9178 \\
$7d_{3/2}$&    -7724&   -981&   298&   10&  -30&   0&   -8427&     -991&  132&   -8604 \\
$7d_{5/2}$&    -7747&   -860&   244&    7&  -28&   0&   -8384&     -856&  109&   -8515 \\
\multicolumn{11}{c}{Ra~II, $Z$ = 88}\\
$7s_{1/2}$&  -75898&  -7529&  2896&  147&  -250&  33&  -80634 & -6692&1152& -81508 \\
$6d_{3/2}$&  -62356&  -8727&  2764&  155&  -398&   0&  -68562 & -8042&1152& -69488 \\
$6d_{5/2}$&  -61592&  -7537&  2202&  114&  -360&   0&  -67174 & -7034& 926& -67947 \\
$7p_{1/2}$&  -56878&  -4182&  1370&  102&  -109&   0&  -59698 & -4027& 587& -60326 \\
$7p_{3/2}$&  -52906&  -3130&  1011&   63&   -90&   0&  -55053 & -3020& 433& -55519 \\
$5f_{5/2}$&  -28660&  -2563&   824&   11&   -63&   0&  -30452 & -4438& 371& -32780 \\
$5f_{7/2}$&  -28705&  -2491&   784&    8&   -61&   0&  -30466 & -4159& 353& -32564 \\
\multicolumn{11}{c}{Ac~III, $Z$ = 89}\\
 $7s_{1/2}$&  -133640&    -9552&   3739&  233&  -357& 58&  -139519&   -8192&  1456&   -140442 \\
 $6d_{3/2}$&  -130697&   -11506&   3639&  296&  -659&  0&  -138927&  -10036&  1479&   -139617 \\
 $6d_{5/2}$&  -128322&   -10002&   2902&  218&  -600&  0&  -135804&   -8884&  1186&   -136401 \\
 $5f_{5/2}$&   -95668&   -26451&   9684&  403& -1785&  0&  -113818&  -23325&  3952&   -116424 \\
 $5f_{7/2}$&   -94161&   -24695&   8837&  289& -1658&  0&  -111387&  -22100&  3607&   -114022 \\
 $7p_{1/2}$&  -106328&    -6202&   2118&  193&  -189&  0&  -110409&   -5688&   874&   -111139 \\
 $7p_{3/2}$&   -98868&    -4745&   1597&  119&  -157&  0&  -102054&   -4380&   659&   -102626 \\
\multicolumn{11}{c}{Th~IV, $Z$ = 90}\\
$5f_{5/2}$&  -206606&  -32100&  11739&   704&  -2747&  0& -229010&    -26327& 4672&  -230304 \\
$5f_{7/2}$&  -203182&  -30549&  10954&   521&  -2616&  0& -224872&    -25252& 4361&  -226168 \\
$6d_{3/2}$&  -211799&  -13258&   4129&   438&   -880&  0& -221370&    -11422& 1663&  -222000 \\
$6d_{5/2}$&  -207574&  -11608&   3300&   326&   -807&  0& -216364&    -10208& 1337&  -216927 \\
$7s_{1/2}$&  -200273&  -11204&   4402&   325&   -458& 89& -207119&     -9455& 1697&  -208075 \\
$7p_{1/2}$&  -165095&   -7991&   2782&   298&   -272&  0& -170278&     -7147& 1125&  -171091 \\
$7p_{3/2}$&  -153572&   -6213&   2124&   184&   -226&  1& -157703&     -5619&  861&  -158372 \\
\multicolumn{11}{c}{Pa~V, $Z$ = 91}\\
$5f_{5/2}$& -336671& -34071&   12323&   953&  -3390&   0&  -360855 &  -27589&  4831&  -361865 \\
$5f_{7/2}$& -331505& -32627&   11592&   713&  -3255&   0&  -355081 &  -26571&  4544&  -356073 \\
$6d_{3/2}$& -303549& -14627&    4452&   586&  -1082&   0&  -314221 &  -12604&  1795&  -314854 \\
$6d_{5/2}$& -297300& -12880&    3556&   438&   -999&   0&  -307185 &  -11354&  1446&  -307768 \\
$7s_{1/2}$& -274949& -12646&    3359&   425&   -557& 126&  -284242 &  -10606&  1348&  -284212 \\
$7p_{1/2}$& -232148&  -9630&    2566&   417&   -356&   0&  -239151 &   -8533&  1040&  -239580 \\
$7p_{3/2}$& -216015&  -7600&    4939&   258&   -297&   1&  -218715 &   -6925&  1900&  -221079 \\
\multicolumn{11}{c}{U~VI, $Z$ = 92}\\
$5f_{5/2}$&  -481613&  -35163&  12554&  1194& -3929&   0&  -506958 & -28393 &  4876&  -507866 \\
$5f_{7/2}$&  -474700&  -33781&  11863&   898& -3789&   0&  -499509 & -27398 &  4604&  -500385 \\
$6d_{3/2}$&  -404911&  -15830&   4675&   742& -1275&   0&  -416599 & -13737 &  1908&  -417273 \\
$6d_{5/2}$&  -396467&  -14000&   3718&   557& -1181&   0&  -407374 & -12472 &  1538&  -408025 \\
$7s_{1/2}$&  -357141&  -13966&   5366&   533&  -655& 173&  -365690 & -11752 &  2079&  -366763 \\
$7p_{1/2}$&  -306870&  -11224&   3742&   548&  -442&   0&  -314245 &  -10135&  1541&  -315358 \\
$7p_{3/2}$&  -285578&   -9134&   2452&   338&  -369&   2&  -292289 &  -10236&  1173&  -294669 \\
\end{tabular}
\end{ruledtabular}
\end{table*}

\begin{table}
\caption{\label{tab1a} The total third-order  $E^{(3)}_\text{
tot}$ and all-order $E^\text{SD}_\text{tot}$ results for Fr-like
ions are compared with experimental
 energies $E_\text{{expt}}$ \protect\cite{expt},
 $\delta E$ = $E_\text{tot}$ - $E_\text{{expt}}$.
 The energies are given relative to the ground state to
 facilitate comparison with experiment. Units: cm$^{-1}$}
\begin{ruledtabular}\begin{tabular}{lrrrrr}
\multicolumn{1}{c}{$nlj$ } & \multicolumn{1}{c}{$E^{(3)}_\text{
tot}$} & \multicolumn{1}{c}{$E^\text{{SD}}_\text{ tot}$} &
\multicolumn{1}{c}{$E_\text{{expt}}$} & \multicolumn{1}{c}{$\delta
E^{(3)}$} &
\multicolumn{1}{c}{$\delta E^\text{{SD}}$} \\
\hline
\multicolumn{5}{c}{Fr~I, $Z$ = 87}\\
$7s_{1/2}$&        0 &      0&      0&     0&     0\\
$7p_{1/2}$&    11683 &  12156&  12237&  -554&   -81\\
$7p_{3/2}$&    13224 &  13838&  13924&  -700&   -86\\
$6d_{3/2}$&    16282 &  16048&  16230&    52&  -192\\
$6d_{5/2}$&    16251 &  16217&  16430&  -179&  -213\\
$8s_{1/2}$&    18893 &  19695&  19733&  -840&   -38\\
$8p_{1/2}$&    22220 &  23037&  23113&  -893&   -76\\
$8p_{3/2}$&    22737 &  23575&  23658&  -921&   -83\\
$7d_{3/2}$&    23418 &  24149&  24245&  -827&   -96\\
$7d_{5/2}$&    23461 &  24238&. 24333&  -872&   -95\\
\multicolumn{5}{c}{Ra~II, $Z$ = 88}\\
$7s_{1/2}$&      0&      0&      0&      0&      0\\
$6d_{3/2}$&  12072&  12020&  12084&    -12&    -64\\
$6d_{5/2}$&  13460&  13561&  13743&   -283&   -182\\
$7p_{1/2}$&  20936&  21182&  21351&   -415&   -169\\
$7p_{3/2}$&  25581&  25989&  26209&   -628&   -220\\
$5f_{5/2}$&  50182&  48728&  48988&   1194&   -260\\
$5f_{7/2}$&  50168&  48944&  49272&    896&   -328\\
\multicolumn{5}{c}{Ac~III, $Z$ = 89}\\
 $7s_{1/2}$&     0&       0&       0&       0&   0\\
 $6d_{3/2}$&   592&     825&     801&    -209&  24\\
 $6d_{5/2}$&  3715&    4041&    4204&    -488&-163\\
 $5f_{5/2}$& 25701&   24018&   23454&    2247& 564\\
 $5f_{7/2}$& 28132&   26420&   26080&    2052& 340\\
 $7p_{1/2}$& 29110&   29303&   29466&    -356&-163\\
 $7p_{3/2}$& 37465&   37816&   38063&    -598&-247\\
\multicolumn{5}{c}{Th~IV, $Z$ = 90}\\
$5f_{5/2}$&       0&       0&      0&     0&      0\\
$5f_{7/2}$&    4138&    4136&   4325&  -187&   -190\\
$6d_{3/2}$&    7640&    8304&   9193& -1553&   -889\\
$6d_{5/2}$&   12646&   13377&  14486& -1841&  -1109\\
$7s_{1/2}$&   21891&   22229&  23131& -1240&   -901\\
$7p_{1/2}$&   58732&   59213&  60239& -1507&  -1026\\
$7p_{3/2}$&   71307&   71932&  73056& -1749&  -1124\\
\multicolumn{5}{c}{U~VI, $Z$ = 92}\\
$5f_{5/2}$&         0&       0&       0&     0&    0\\
$5f_{7/2}$&      7449&    7481&    7609&  -160& -128\\
$6d_{3/2}$&     90359&   90593&   91000&  -641& -407\\
$6d_{5/2}$&     99584&   99841&  100510&  -926& -669\\
$7s_{1/2}$&    141268&  141103&  141447&  -179& -344\\
$7p_{1/2}$&    192713&  191989&  193340&  -627&  -832\\
$7p_{3/2}$&    214669&  211747&  215886& -1217& -2689\\
\end{tabular}
\end{ruledtabular}
\end{table}

\section{Third-order and all-order MBPT calculations  of energies}

We  start from the ``no-pair'' Hamiltonian  \cite{sucher}
\begin{equation}\label{eq1}
H=H_{0}+V_{I}\,,
\end{equation}
where $H_{0}$ and $V_{I}$ can be written in a second-quantized
form as
\begin{equation}\label{eq2}
H_{0}=\sum_{i}\varepsilon _{i}a_{i}^{\dagger}a_{i}\,,
\end{equation}
\begin{equation}\label{eq3}
V_{I}=\frac{1}{2}\sum_{ijkl}g_{ijkl}\
a_{i}^{\dagger}a_{j}^{\dagger}a_{l}a_{k}\, .
\end{equation}
Negative-energy (positron) states are excluded from the sums. The
quantities $\varepsilon _{i}$ are eigenvalues of the one-electron
Dirac-Fock  equations with a frozen core and  $g_{ijkl}$ is
a two-particle Coulomb matrix element. Our calculations start
from a $V^{N-1}$ DF potential for a closed-subshell radon-like ion.

The all-order single-double (SD) method was discussed previously in
Refs.~\cite{blundell-li,Liu,blundell-cs,safr-na,safr-alk,gold}.
Briefly, we represent the wave function $\Psi _{v}$ of an atom
with one valence electron atom as $\Psi _{v}\cong \Psi _{v}^{\rm
SD}$ with
\begin{eqnarray}\label{eq8}
\lefteqn{ \Psi _{v}^{\rm SD} =\left[ 1+\sum_{ma}\rho _{ma}a_{m}^\dagger a_{a}+\frac{1}{2}%
\sum_{mnab}\rho _{mnab}a_{m}^\dagger a_{n}^\dagger a_{b}a_{a}\right. }
\hspace{1.5em} \nonumber
\\ &&\left. +\sum_{m\neq v}\rho
_{mv}a_{m}^\dagger a_{v}+\sum_{mna}\rho
_{mnva}a_{m}^\dagger a_{n}^\dagger a_{a}a_{v}\right]\!\! \Phi _{v},
\end{eqnarray}
where $\Phi _{v}$ is the lowest-order atomic wave function, which
is taken to be the frozen-core DF wave function of a state $v$.
Substituting the wave function $\Psi _{v}^{\rm SD}$ into the
many-body Schr\"{o}dinger equation, with Hamiltonian given by the
Eqs.~(\ref{eq1}--\ref{eq3}), one obtains the coupled
equations for the single- and double-excitation coefficients $\rho _{mv}$,  $%
\rho _{ma}$, $\rho _{mnva}$, and  $\rho _{mnab}$. The coupled
equations  for the excitation coefficients are solved iteratively.
We use the resulting excitation coefficients to evaluate multipole
matrix elements  and hyperfine constants. This method includes
contribution of important classes of MBPT corrections to all orders.

The SD valence $E_{v}^{\rm SD}$ energy does not include a all
third-order MBPT corrections. The missing part of the third-order
contribution, $E_{\rm extra}^{(3)}$, is written out in
Ref.~\cite{safr-na}  and must be calculated separately. We use
our third-order energy code to separate out $E_{\rm extra}^{(3)}$
and add it to the $E_{v}^{\rm SD}$. For notational simplicity,
we drop the index $v$ in the designations in the text and tables
below.

\begin{figure*}[tbp]
\centerline{\includegraphics[scale=0.35]{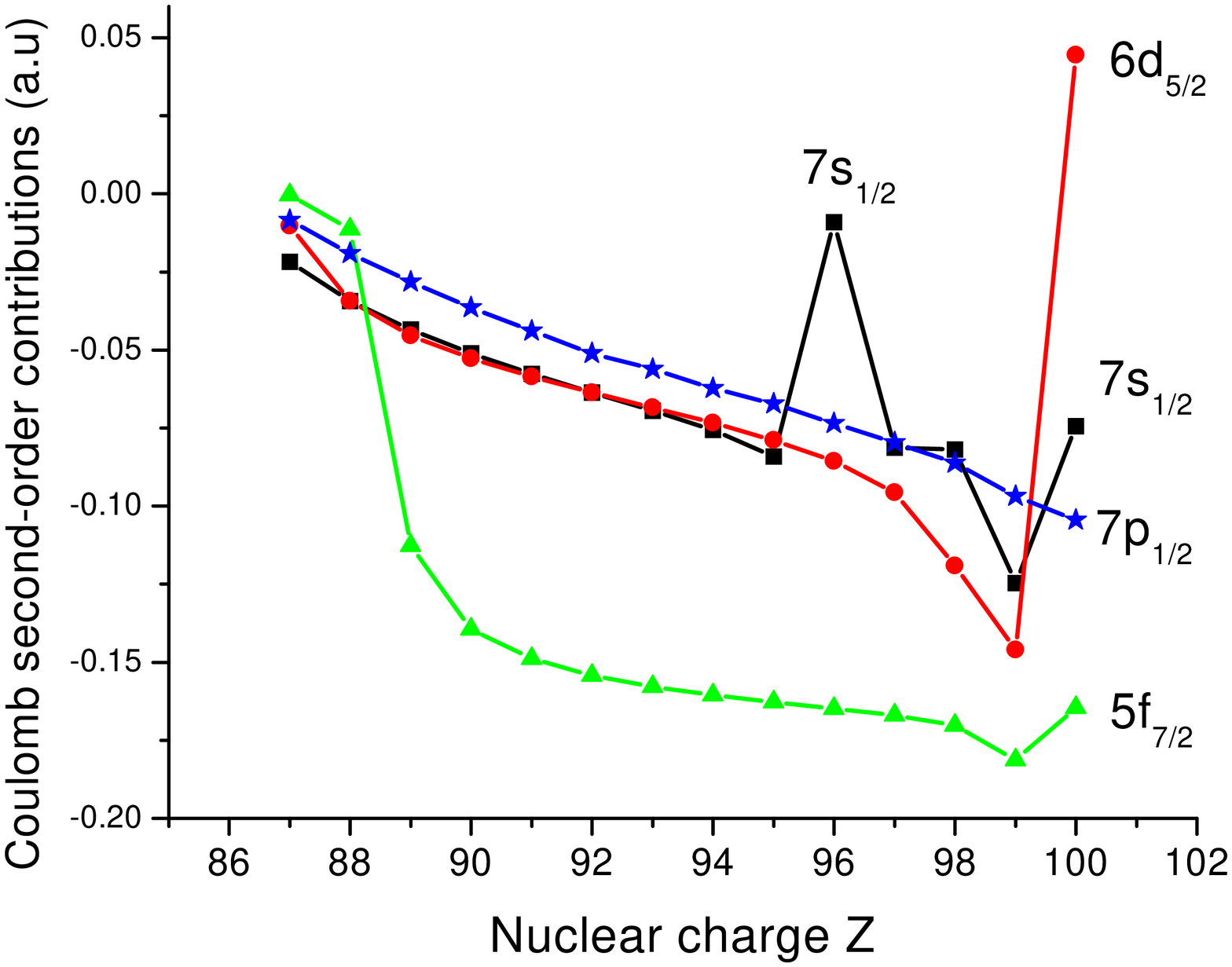}
            \includegraphics[scale=0.35]{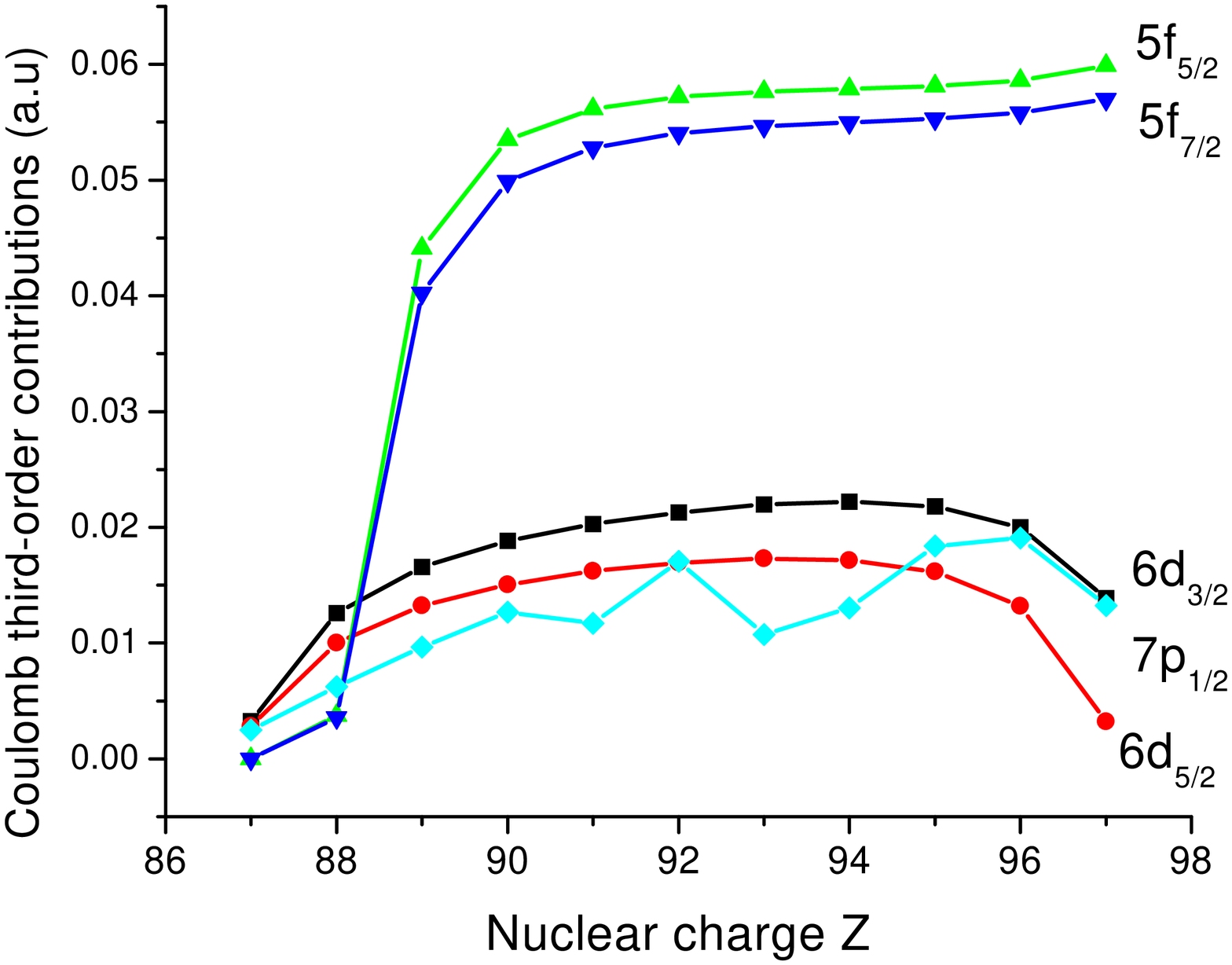}}
\caption{Coulomb second and third-order energies as functions of
$Z$ for the $5f_{j}$, $6d_{j}$,  $7s_{1/2}$, and  $7p_{1/2}$
states
 in Fr-like ions.}
\label{fig-e2}
\end{figure*}

Results of our  energy calculations for low-lying  states of
Fr~I--U~VI are summarized in
Table~\ref{tab1}. Columns 2--7 of Table~\ref{tab1} give the
lowest-order DF energies $E^{(0)}$, second- and third-order
Coulomb correlation energies, $E^{(2)}$ and $E^{(3)}$,
 first-order Breit contribution $B^{(1)}$,  second-order
Coulomb-Breit $B^{(2)}$ corrections, and the Lamb shift
contribution, $E_{\rm LS}$. The sum of these six
contributions is our final third-order MBPT result
$E^{(3)}_{\rm tot}$ listed in the eighth column.
First-order Breit energies (column
$B^{(1)}$ of Table~\ref{tab1}) include retardation, whereas the
second-order Coulomb-Breit energies (column $B^{(2)}$ of
Table~\ref{tab1}) are evaluated using the unretarded Breit
operator. We list all-order SD energies in the column labelled
$E^\text{{SD}}$ and the part of the third-order energies omitted
in the SD calculation in column $E^{(3)}_\text{{extra}}$. We note
that $E^\text{{SD}}$ includes $E^{(2)}$, part of $E^{(3)}$, and dominant
higher-order corrections. The
sum of the six terms $E^{(0)}$, $E^\text{{SD}}$,
$E^{(3)}_\text{{extra}}$, $B^{(1)}$,  $B^{(2)}$, and  $E_{\rm LS}$
gives the final all-order results $E^\text{{SD}}_{\rm tot}$ listed
in the eleventh column of the table.

As expected, the largest correlation contribution to the valence
energy  comes from the second-order term, $E^{(2)}$.
 This term  is  simple to calculate in comparison with
$E^{(3)}$ and $E^{\rm SD}$ terms. Thus, we calculate the $E^{(2)}$
term with higher numerical accuracy than $E^{(3)}$ and
$E^{\rm SD}$. The second-order  energy $E^{(2)}$ includes partial
waves up to $l_{\text{max}}=8$ and is extrapolated to account for
contributions from higher partial waves (see, for example,
Refs.~\cite{be-en,be3-en}). As an example of the convergence of
$E^{(2)}$ with the number of partial waves $l$, we consider the
$5f_{5/2}$ state in  U~VI. Calculations of $E^{(2)}$ with
$l_{\text{max}}$ = 6 and  8 yield
 $E^{(2)}(5f_{5/2})$ = -33598 and
-347559$~$cm$^{-1}$, respectively. Extrapolation of these
calculations yields -35163 and -35227~cm$^{-1}$, respectively.
Therefore, we estimate the numerical uncertainty of
$E^{(2)}(5f_{5/2})$ to be  64~cm$^{-1}$.   This is the largest
contribution from the higher partial waves, since  the numerical
uncertainty of $E^{(2)}(6d)$ is equal to 34~cm$^{-1}$, and the
numerical uncertainty of $E^{(2)}(7s)$ is equal to
1~cm$^{-1}$. The numerical uncertainty of the second-order energy
calculation for all other states ranges from 1~cm$^{-1}$ to
5~cm$^{-1}$. We use $l_{\text{max}}$ = 6  in our third-order and
all-order calculations, owing to the complexity of
these calculations.  Therefore, we use our high-precision
calculation of $E^{(2)}$ described above to account for the
contributions of the higher partial waves, i.e. we replace
$E^{(2)}$[$l_{\text{max}}$ = 6] with the final
high-precision second-order value $E^{(2)}_{\rm final}$.  The
contribution $E^{(3)}_\text{{extra}}$ given
 in Table~\ref{tab1} accounts for that part of the
third-order MBPT correction not included in the SD
energy. The values of $E^{(3)}_\text{{extra}}$ are quite large and
including this term is important.

 We find that the correlation corrections to
energies  are  especially large for $5f$ states.  For
example, $E^{(2)}$ is about 15\% of $E^{(0)}$ and $E^{(3)}$ is
about 36\% of $E^{(2)}$ for $5f$ states. Despite the
evident slow convergence of the perturbation theory expansion, the
$5f$ energy from the third-order MBPT calculation is within
0.9\% of the measured energy. The correlation corrections are so
large for the $5f$ states that inclusion of correlation leads
to the different ordering of states using the DF and
$E^{(3)}_\text{ tot}$ energies. If we consider only DF energies
the $6d_{3/2}$ appears to be the ground state for Th~IV, but the full
correlation shows that the $5f_{5/2}$ is the ground state. The
correlation corrections are much smaller for all other states; the
ratios of $E^{(0)}$ and $E^{(2)}$ are equal to 6\%, 5\%, and 2\%
for the $6d$, $7s$, and $10s$ states,
respectively.

 The third-order and all-order results are compared with
experimental values in Table~\ref{tab1a}. The energies are given
relative to the ground state to facilitate comparison with
experiment.
 Experimental
energies for Fr~I, Ac~III, Th~IV, and U~VI are taken from
\cite{expt} and  energies of Fr-like Ra  are taken from the NIST
compilation \cite{web-nist}.  Differences of our
third-order and all-order
 calculations with
experimental data, $\delta E^{(3)}=E^{(3)}_{\rm
tot}-E_\text{{expt}}$ and $\delta E^\text{SD}=E^\text{{SD}}_{\rm
tot}-E_\text{{expt}}$, respectively, are given in the two final
columns of Table~\ref{tab1a}. In general, the SD results  agree
better with the experimental values than the third-order MBPT
values. Exceptions are the cases where the third-order
fortuitously give results that are close to experimental values.
Comparison of results from two last columns of
Table~\ref{tab1a} shows that the ratio of $\delta E^{(3)}$ and
$\delta E^\text{{SD}}$ is about three for the $5f$ states. As
expected, including correlation to all orders led to significant
improvement of the results. Better agreement of
all-order energies with experiment demonstrates the importance of
the higher-order correlation contributions.

Below, we describe a few  numerical details of the calculation. We
use the B-spline method described \cite{Bspline} to generate a complete
set of
basis DF wave functions for use in the evaluation of the MBPT
expressions. We use 50 splines of order $k=8$ for each angular
momentum. The basis orbitals are constrained to a spherical cavity
of radius  $R$ = 90--30 a.u for Fr~I--U~VI. The cavity radius is
chosen large enough to accommodate all $nl_j$  orbitals considered
here and small enough that 50 splines can approximate inner-shell
DF wave functions with good precision. We use 40 out 50 basis
orbitals for each partial wave in our third-order and all-order
energy calculations, since contributions from the  highest-energy
orbitals are negligible.

\begin{figure}[tbp]
\centerline{\includegraphics[scale=0.35]{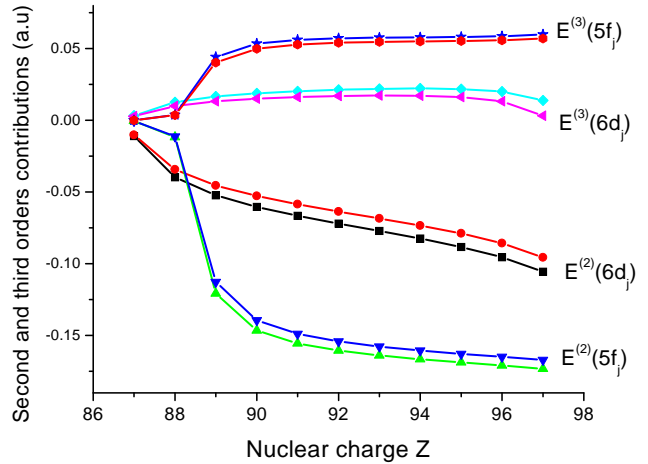}}
\caption{Coulomb second- and third-order energies as functions of
$Z$ for the $5f_{j}$ and $6d_{j}$ states
 in Fr-like ions.}
\label{fig-e3}
\end{figure}

\begin{figure}[tbp]
\centerline{\includegraphics[scale=0.35]{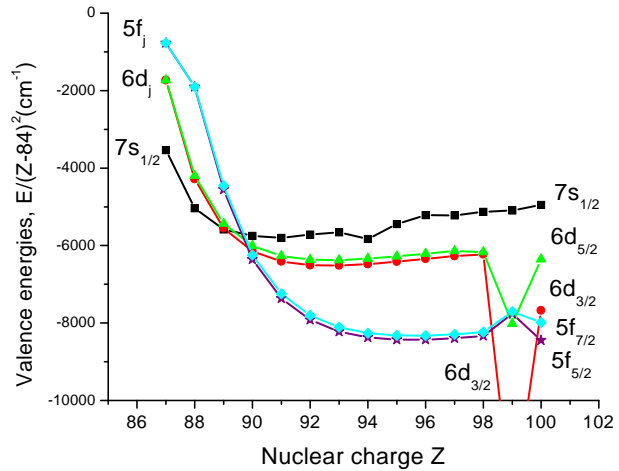}}
\caption{Valence removal energies ($E/(Z-84)^2$ in cm$^{-1}$)
 as functions of $Z$ for the $5f_j$, $6d_j$, and $7s_{1/2}$
states in Fr-like ions. }
 \label{fig-en}
\end{figure}

\subsection{$Z$ dependence of  energies in Fr-like ions}
In Fig.~\ref{fig-e2}, we illustrate the $Z$-dependence of the
second and third-order energy corrections $E^{(n)}$ for the valence
$7s$, $7p$, $6d$, and $5f$ states of Fr-like ions.
The second-order energy $E^{(2)}$ is a smooth function of $Z$ for
the $7p_{1/2}$ and $5f_{7/2}$ states, but exhibits a few sharp
features for the $7s_{1/2}$ and $6d_{5/2}$ states.
 These very strong irregularities occur for the $7s_{1/2}$
  state $Z$=96, 99 and
 $6d_{5/2}$ state for
$Z$=99 and are explained  by accidentally small energy
denominators in the MBPT expressions for the correlation corrections
to the energy. The third-order energy $E^{(3)}$ is a smooth
function of $Z$ for the $6d$ and $5f$ states, but exhibits
a similar sharp features for the $7p_{1/2}$
 state.  Most of the sharp features for the
 $6d_{j}$ and $7p_{j}$ states occur at very high values of $Z$, $Z>$ 97.
 Comparison of the $E^{(2)}$ and $E^{(3)}$ corrections for
 the $6d$ and
$5f$ states is illustrated by Fig.~\ref{fig-e3}.
This figure shows also the smooth dependence of second- and
third-order corrections to the $6d_{5/2} - 6d_{3/2}$ and $5f_{7/2}
- 5f_{5/2} $ fine-structure intervals.

The total $E^{(3)}_\text{ tot}$ energies divided by $(Z-84)^2$ in
Fr-like ions are shown in Fig.~\ref{fig-en}. We plot the five
energy levels for the $7s_{1/2}$, $6d_{j}$, and $5f_{j}$ states.
The  $Z$-dependence of $E^{(3)}_\text{ tot}$ is smooth  up to very
high $Z$.  Irregularities observed in  Fig.~\ref{fig-en} for
high $Z$ are explained by vanishing energy denominators in MBPT
expressions for correlation corrections.

\begin{table*}
\caption{\label{tab-dip} Reduced electric-dipole matrix elements
in Fr-like ions
 calculated to first, second, third, and all
orders of MBPT.}
\begin{ruledtabular}
\begin{tabular}{llrrrrllrrrr}
\multicolumn{2}{c}{Transition}& \multicolumn{1}{c}{$Z^{({\rm
DF})}$ }& \multicolumn{1}{c}{$Z^{({\rm DF}+2)}$ }&
\multicolumn{1}{c}{$Z^{({\rm DF}+2+3)}$ }&
\multicolumn{1}{c}{$Z^\text{{(SD)}}$ }&
\multicolumn{2}{c}{Transition}& \multicolumn{1}{c}{$Z^{({\rm
DF})}$ }& \multicolumn{1}{c}{$Z^{({\rm DF}+2)}$ }&
\multicolumn{1}{c}{$Z^{({\rm DF}+2+3)}$ }&
\multicolumn{1}{c}{$Z^\text{{(SD)}}$ }\\
\hline \multicolumn{6}{c}{Fr~I, $Z$ = 87}&
\multicolumn{6}{c}{Th~IV, $Z$ = 90}\\
$7s_{1/2}$&$7p_{1/2}$&   5.1438&  4.7412&  4.1362&  4.2641& $7s_{1/2}$&$ 7p_{1/2}$& 2.8994& 2.3748& 2.3669& 2.4196\\
$7s_{1/2}$&$7p_{3/2}$&   7.0903&  6.6001&  5.6451&  5.8619& $7s_{1/2}$&$ 7p_{3/2}$& 3.9933& 3.3399& 3.2930& 3.3677\\
$6d_{3/2}$&$7p_{1/2}$&   9.2216&  8.7781&  7.1945&  7.0276& $6d_{3/2}$&$ 7p_{1/2}$& 2.5465& 2.1374& 2.0723& 2.1220\\
$6d_{3/2}$&$7p_{3/2}$&   4.2832&  4.1100&  3.3108&  3.2217& $6d_{3/2}$&$ 7p_{3/2}$& 0.9963& 0.8784& 0.8270& 0.8488\\
$6d_{5/2}$&$7p_{3/2}$&  12.8041& 12.2858& 10.1204&  9.9456& $6d_{5/2}$&$ 7p_{3/2}$& 3.1975& 2.8348& 2.7006& 2.7549\\
$6d_{3/2}$&$5f_{5/2}$&  11.4529& 11.3002&  6.7352&  6.9611& $6d_{3/2}$&$ 5f_{5/2}$& 2.4281& 1.4501& 1.3367& 1.5295\\
$6d_{5/2}$&$5f_{5/2}$&   3.0143&  2.9714&  1.9006&  1.9515& $6d_{5/2}$&$ 5f_{5/2}$& 0.6391& 0.4070& 0.3624& 0.4116\\
$6d_{5/2}$&$5f_{7/2}$&  13.4863& 13.2942&  8.5062&  8.7332& $6d_{5/2}$&$ 5f_{7/2}$& 2.9557& 1.8844& 1.7032& 1.9190\\
\multicolumn{6}{c}{Ra~II, $Z$ = 88}&
\multicolumn{6}{c}{Pa~V, $Z$ = 91}\\
$7s_{1/2}$&$7p_{1/2}$&   3.8766&  3.3758&  3.1793&  3.2545& $7s_{1/2}$&$ 7p_{1/2}$& 2.6267& 2.1111& 2.1365& 2.1822\\
$7s_{1/2}$&$7p_{3/2}$&   5.3395&  4.7241&  4.3881&  4.5106& $7s_{1/2}$&$ 7p_{3/2}$& 3.6173& 2.9720& 2.9750& 3.0327\\
$6d_{3/2}$&$7p_{1/2}$&   4.4462&  3.9314&  3.3975&  3.5659& $6d_{3/2}$&$ 7p_{1/2}$& 2.1785& 1.8079& 1.7817& 1.8152\\
$6d_{3/2}$&$7p_{3/2}$&   1.8815&  1.7070&  1.4262&  1.5117& $6d_{3/2}$&$ 7p_{3/2}$& 0.8319& 0.7327& 0.7004& 0.7119\\
$6d_{5/2}$&$7p_{3/2}$&   5.8616&  5.3340&  4.6240&  4.8232& $6d_{5/2}$&$ 7p_{3/2}$& 2.6895& 2.3817& 2.2936& 2.3185\\
$6d_{3/2}$&$5f_{5/2}$&   5.3548&  4.6786&  4.5967&  4.4491& $6d_{3/2}$&$ 5f_{5/2}$& 1.9275& 1.0904& 1.0864& 1.2092\\
$6d_{5/2}$&$5f_{5/2}$&   1.4797&  1.3062&  1.3088&  1.2465& $6d_{5/2}$&$ 5f_{5/2}$& 0.5023& 0.3061& 0.2947& 0.3241\\
$6d_{5/2}$&$5f_{7/2}$&   6.6382&  5.8609&  5.8347&  5.6357& $6d_{5/2}$&$ 5f_{7/2}$& 2.3242& 1.4102& 1.3673& 1.5020\\
\multicolumn{6}{c}{Ac~III, $Z$ = 89}&
\multicolumn{6}{c}{U~VI, $Z$ = 92}\\
$7s_{1/2}$&$7p_{1/2}$&   3.2787&  2.7544&  2.6859&  2.7463& $7s_{1/2}$&$ 7p_{1/2}$& 2.4165& 1.9138& 1.9562& 1.9599\\
$7s_{1/2}$&$7p_{3/2}$&   4.5157&  3.8665&  3.7271&  3.8176& $7s_{1/2}$&$ 7p_{3/2}$& 3.3271& 2.6952& 2.7168& 2.2449\\
$6d_{3/2}$&$7p_{1/2}$&   3.1529&  2.6942&  2.5241&  2.6048& $6d_{3/2}$&$ 7p_{1/2}$& 1.9249& 1.5837& 1.5688& 1.5613\\
$6d_{3/2}$&$7p_{3/2}$&   1.2726&  1.1300&  1.0275&  1.0662& $6d_{3/2}$&$ 7p_{3/2}$& 0.7202& 0.6353& 0.6054& 0.4963\\
$6d_{5/2}$&$7p_{3/2}$&   4.0423&  3.6067&  3.3438&  3.4383& $6d_{5/2}$&$ 7p_{3/2}$& 2.3417& 2.0760& 1.9887& 1.6018\\
$6d_{3/2}$&$5f_{5/2}$&   3.5764&  2.3879&  1.7460&  2.1624& $6d_{3/2}$&$ 5f_{5/2}$& 1.6274& 0.8798& 0.9034& 0.9974\\
$6d_{5/2}$&$5f_{5/2}$&   0.9577&  0.6708&  0.4760&  0.5877& $6d_{5/2}$&$ 5f_{5/2}$& 0.4209& 0.2479& 0.2470& 0.2674\\
$6d_{5/2}$&$5f_{7/2}$&   4.4148&  3.1076&  2.2956&  2.7539& $6d_{5/2}$&$ 5f_{7/2}$& 1.9492& 1.1353& 1.1293& 1.2293\\
  \end{tabular}
\end{ruledtabular}
\end{table*}

\begin{table*}
\caption{\label{tab-LV} Comparison of length [L] and velocity [V]
results for reduced electric-dipole matrix elements in first,
second, and third orders of perturbation theory in  Fr-like
systems with $Z$ = 87--92. }
\begin{ruledtabular}
\begin{tabular}{lrrrrrrrrrrrr}
\multicolumn{1}{c}{Ion}& \multicolumn{2}{c}{$Z^{({\rm DF})}$ }&
\multicolumn{2}{c}{$Z^{({\rm DF}+2)}$ }&
\multicolumn{2}{c}{$Z^{({\rm DF}+2+3)}$}&
\multicolumn{2}{c}{$Z^{({\rm DF})}$ }&
\multicolumn{2}{c}{$Z^{({\rm DF}+2)}$ }&
\multicolumn{2}{c}{$Z^{({\rm DF}+2+3)}$}\\
\multicolumn{1}{c}{ }& \multicolumn{1}{c}{ $L$}&
\multicolumn{1}{c}{ $V$}& \multicolumn{1}{c}{ $L$}&
\multicolumn{1}{c}{ $V$}& \multicolumn{1}{c}{ $L$}&
\multicolumn{1}{c}{ $V$}& \multicolumn{1}{c}{ $L$}&
\multicolumn{1}{c}{ $V$}& \multicolumn{1}{c}{ $L$}&
\multicolumn{1}{c}{ $V$}& \multicolumn{1}{c}{ $L$}&
\multicolumn{1}{c}{ $V$}\\
\hline \multicolumn{6}{c}{$6d_{3/2} - 5f_{5/2}$}&
 \multicolumn{6}{c}{$6d_{5/2} - 5f_{7/2}$}\\
  87& 11.4529&  11.4002&  11.2693&  11.2693&   7.8982&   7.8972&  13.4863&  13.4022&  13.2703&  13.2703&   9.6464&   9.6455 \\
  88&  5.3549&   5.1362&   4.7603&   4.7603&   4.4885&   4.4839&   6.6382&   6.3283&   5.9720&   5.9720&   5.7071&   5.7056 \\
  89&  3.5764&   3.1632&   2.6134&   2.6134&   0.5254&   0.4982&   4.4148&   3.8730&   3.3832&   3.3831&   0.8484&   0.8576 \\
  90&  2.4281&   0.7698&   1.6597&   1.6595&   2.8562&   2.9000&   2.9557&   0.5609&   2.1257&   2.1255&   3.3454&   3.3138 \\
  91&  1.9275&   2.0349&   1.2702&   1.2702&   1.4017&   1.4122&   2.3242&   2.4373&   1.6097&   1.6097&   1.7219&   1.7100 \\
  92&  1.6274&   1.6199&   1.0312&   1.0312&   1.0559&   1.0613&   1.9492&   1.9305&   1.2978&   1.2978&   1.3024&   1.2943 \\
\multicolumn{6}{c}{$6d_{3/2} - 7p_{1/2}$}&
\multicolumn{6}{c}{$7s_{3/2} - 7p_{3/2}$}\\
  87& 9.2216&   9.9449&   8.7906&   8.7906&   6.8407&   6.8365&   7.0903&   6.6425&   6.6268&   6.6268&   5.9488&   5.9486 \\
  88& 4.4462&   2.0213&   3.9815&   3.9816&   3.8547&   3.8589&   5.3395&   4.9435&   4.7968&   4.7969&   4.5150&   4.5148 \\
  89& 3.1529&   2.3683&   2.7533&   2.7533&   2.6221&   2.6238&   4.5158&   4.1785&   3.9641&   3.9641&   3.8018&   3.8016 \\
  90& 2.5465&   2.0531&   2.1960&   2.1961&   2.1074&   2.1084&   3.9933&   3.7030&   3.4515&   3.4516&   3.3434&   3.3432 \\
  91& 2.1785&   1.8158&   1.8631&   1.8631&   1.7957&   1.7964&   3.6173&   3.3638&   3.0915&   3.0915&   3.0121&   3.0119 \\
  92& 1.9249&   1.6384&   1.6343&   1.6344&   1.5743&   1.5748&   3.3271&   3.1029&   2.8190&   2.8190&   2.7460&   2.7458 \\
  \end{tabular}
\end{ruledtabular}
\end{table*}

\begin{table*}
\caption{\label{tab-com1} Wavelengths $\lambda$ (\AA), weighted
transition rates $gA$ (s$^{-1}$)   and oscillator strengths $gf$
in Ac~III, Th~IV, and U~VI. The SD data
 ($gA^{\mathrm{(SD)}}$ and $gf^{\mathrm{(SD)}}$) are compared with
  theoretical ($gA^{\mathrm{(RHF)}}$ and $gf^{\mathrm{(RHF)}}$)
  values
  given in Ref.~\protect\cite{osc-ra}.  Numbers in brackets represent powers of 10. }
\begin{ruledtabular}
\begin{tabular}{llrllll}
\multicolumn{1}{c}{Lower}& \multicolumn{1}{c}{Upper}&
\multicolumn{1}{c}{$\lambda^\text{{(expt)}}$}&
\multicolumn{1}{c}{$gA^{\mathrm{(SD)}}$ }&
\multicolumn{1}{c}{$gA^{\mathrm{(RHF)}}$ }&
\multicolumn{1}{c}{$gf^{\mathrm{(SD)}}$ }&
\multicolumn{1}{c}{$gf^{\mathrm{(RHF)}}$ }\\
\hline
 \multicolumn{7}{c}{Ac~III, $Z$ = 89}\\
 $7s_{1/2}$&$7p_{3/2}$&    2626.44&   1.63[9]&   1.55[9] &  1.69[+0]&   1.6[+0]\\
 $6d_{3/2}$&$7p_{3/2}$&    2682.90&   1.19[8]&   1.23[8] &  1.29[-1]&   1.3[-1]\\
 $6d_{5/2}$&$7p_{3/2}$&    2952.55&   9.30[8]&   8.23[8] &  1.22[+0]&   1.1[+0]\\
 $7s_{1/2}$&$7p_{1/2}$&    3392.78&   3.91[8]&   3.59[8] &  6.75[-1]&   6.2[-1]\\
 $6d_{3/2}$&$7p_{1/2}$&    3487.59&   3.24[8]&   2.75[8] &  5.91[-1]&   5.0[-1]\\
 $6d_{3/2}$&$5f_{5/2}$&    4413.09&   1.10[8]&   2.34[8] &  3.22[-1]&   6.9[-1]\\
 $6d_{5/2}$&$5f_{7/2}$&    4569.97&   1.61[8]&   3.02[8] &  5.04[-1]&   9.5[-1]\\
 $6d_{5/2}$&$5f_{5/2}$&    5193.21&   4.99[6]&   1.03[6] &  2.02[-2]&   4.2[-2]\\
  \multicolumn{7}{c}{Th~IV, $Z$ = 90}\\
 $6d_{3/2}$&$7p_{3/2}$&    1565.86&   3.80[8]&   4.08[8] &  1.40[-1]&   1.5[-1]\\
 $6d_{5/2}$&$7p_{3/2}$&    1707.37&   3.09[9]&   2.83[9] &  1.35[+0]&   1.2[+0]\\
 $6d_{3/2}$&$7p_{1/2}$&    1959.02&   1.21[9]&   1.04[9] &  6.98[-1]&   6.0[-1]\\
 $7s_{1/2}$&$7p_{3/2}$&    2003.00&   2.86[9]&   2.70[9] &  1.72[+0]&   1.6[+0]\\
 $7s_{1/2}$&$7p_{1/2}$&    2694.81&   6.06[8]&   5.55[8] &  6.60[-1]&   6.0[-1]\\
 $5f_{5/2}$&$6d_{5/2}$&    6903.05&   1.04[6]&   2.22[6] &  7.45[-3]&   1.6[-2]\\
 $5f_{7/2}$&$6d_{5/2}$&    9841.54&   7.83[6]&   1.53[7] &  1.14[-1]&   2.2[-1]\\
 $5f_{5/2}$&$6d_{3/2}$&   10877.55&   3.68[6]&   7.93[6] &  6.53[-2]&   1.4[-1]\\
   \multicolumn{7}{c}{U~VI, $Z$ = 92}\\
 $6d_{3/2}$&$7p_{3/2}$&    800.729&   9.72[8]&   1.72[9] &  9.34[-2]&   1.7[-1]\\
 $6d_{5/2}$&$7p_{3/2}$&    866.737&   7.98[9]&   1.26[10]&  8.99[-1]&   1.4[+0]\\
 $6d_{3/2}$&$7p_{1/2}$&    977.129&   5.29[9]&   4.87[9] &  7.58[-1]&   6.9[-1]\\
 $5f_{5/2}$&$6d_{5/2}$&    994.921&   1.47[8]&   3.51[8] &  2.18[-2]&   5.2[-2]\\
 $5f_{7/2}$&$6d_{5/2}$&    1076.40&   2.46[9]&   5.55[9] &  4.26[-1]&   9.5[-1]\\
 $5f_{5/2}$&$6d_{3/2}$&    1098.91&   1.52[9]&   3.64[9] &  2.75[-1]&   6.6[-1]\\
 $7s_{1/2}$&$7p_{3/2}$&    1343.39&   4.21[9]&   6.08[9] &  1.14[+0]&   1.7[+0]\\
 $7s_{1/2}$&$7p_{1/2}$&    1927.05&   1.09[9]&   1.03[9] &  6.05[-1]&   5.8[-1]\\
  \end{tabular}
\end{ruledtabular}
\end{table*}

\begin{table*}
\caption{\label{tab-com2} Wavelengths $\lambda$ (\AA), weighted
transition rates $gA$ (s$^{-1}$)  and oscillator strengths $gf$ in
Fr~I and Ra~II. The SD data
 ($gA^{\mathrm{(SD)}}$ and $gf^{\mathrm{(SD)}}$) are compared with
  theoretical ($gA^{\mathrm{(RHF)}}$,  $gf^{\mathrm{(RHF)}}$)
  values
  given in Ref.~\protect\cite{biemont-fr,osc-ra} and
  theoretical ($gA^{\mathrm{(MBPT)}}$,  $gf^{\mathrm{(MBPT)}}$)
  values
  given in Ref.~\protect\cite{dzuba-01}.  Numbers in brackets represent powers of 10. }
\begin{ruledtabular}
\begin{tabular}{llrllllll}
\multicolumn{1}{c}{Lower}& \multicolumn{1}{c}{Upper}&
\multicolumn{1}{c}{$\lambda^\text{{(expt)}}$}&
\multicolumn{1}{c}{$gA^{\mathrm{(SD)}}$ }&
\multicolumn{1}{c}{$gA^{\mathrm{(MBPT)}}$ }&
\multicolumn{1}{c}{$gA^{\mathrm{(RHF)}}$ }&
\multicolumn{1}{c}{$gf^{\mathrm{(SD)}}$ }&
\multicolumn{1}{c}{$gf^{\mathrm{(MBPT)}}$ }&
\multicolumn{1}{c}{$gf^{\mathrm{(RHF)}}$ }\\
\hline
    \multicolumn{9}{c}{Fr~I, $Z$ = 87}\\
 $7s_{1/2}$&$7p_{3/2}$&  7181.84&  1.88[8]&       &          &    1.45[+0]&    1.46[+0]&    1.5[+0]\\
 $7s_{1/2}$&$7p_{1/2}$&  8171.66&  6.75[7]&       &          &    6.76[-1]&    6.84[-1]&    6.5[-1]\\
 $7p_{1/2}$&$7d_{3/2}$&  8328.18&  4.39[7]&       &          &    4.57[-1]&    4.40[-1]&    6.9[-1]\\
 $7p_{3/2}$&$7d_{5/2}$&  9606.79&  8.83[7]&       &          &    1.22[+0]&    1.18[+0]&    1.1[+0]\\
 $7p_{3/2}$&$7d_{3/2}$&  9689.14&  1.08[7]&       &          &    1.52[-1]&    1.49[-1]&    1.2[-1]\\
 $7p_{1/2}$&$8s_{1/2}$&  13342.0&  1.49[7]&       &          &    3.99[-1]&            &           \\
 $7p_{3/2}$&$8s_{1/2}$&  17216.1&  2.18[7]&       &          &    9.70[-1]&            &           \\
   \multicolumn{9}{c}{Ra~II, $Z$ = 88}\\
 $6d_{3/2}$&$5f_{5/2}$&  2708.96&  2.02[9]&        &  1.59[9]&    2.22[+0]&            &    1.7[+0]\\
 $6d_{5/2}$&$5f_{7/2}$&  2813.76&  2.89[9]&        &  2.01[9]&    3.43[+0]&            &    2.4[+0]\\
 $6d_{5/2}$&$5f_{5/2}$&  2836.46&  1.38[8]&        &  9.80[7]&    1.66[-1]&            &    1.2[-1]\\
 $7s_{1/2}$&$7p_{3/2}$&  3814.42&  7.42[8]&  7.30[8]& 7.14[8]&    1.62[+0]&    1.59[+0]&    1.5[+0]\\
 $7s_{1/2}$&$7p_{1/2}$&  4682.28&  2.09[8]&  2.05[8]& 1.93[8]&    6.87[-1]&    6.73[-1]&    6.3[-1]\\
 $6d_{3/2}$&$7p_{3/2}$&  7077.95&  1.30[7]&  1.29[7]& 1.27[7]&    9.80[-2]&    9.70[-2]&    9.5[-2]\\
 $6d_{5/2}$&$7p_{3/2}$&  8019.70&  9.13[7]&  9.10[7]& 7.85[7]&    8.81[-1]&    8.78[-1]&    7.6[-1]\\
 $6d_{3/2}$&$7p_{1/2}$&  10788.2&  2.05[7]&  2.03[7]& 1.79[7]&    3.58[-1]&    3.54[-1]&    3.1[-1]\\
  \end{tabular}
\end{ruledtabular}
\end{table*}

\begin{table}
\caption{\label{tab-life} Lifetimes ${\tau}$ in ns
 for the $nl$  levels
 for  neutral francium. Our SD results are compared with experimental
 measurements presented in Ref.~\protect\cite{fr-98} for the
 $7p_j$ levels, in Ref.~\protect\cite{fr-00} for the
 $7d_j$ levels, and in Ref.~\protect\cite{fr-05} for the
 $8s_{1/2}$ level.}
\begin{ruledtabular}
\begin{tabular}{lll}
\multicolumn{1}{c}{Level}              &
\multicolumn{1}{c}{$\tau^{(\rm SD)}$}  &
\multicolumn{1}{c}{$\tau^{(\rm expt)}$} \\\hline
   $7p_{1/2}$&     29.62  & 29.45$\pm$0.11 \\
   $7p_{3/2}$&     21.28  & 21.02$\pm$0.11  \\
   $7d_{3/2}$&     73.08  &  73.6$\pm$0.3 \\
   $7d_{5/2}$&     67.93  & 67.7$\pm$2.9  \\
   $8s_{1/2}$&     54.36  & 53.30$\pm$0.44  \\
\end{tabular}
\end{ruledtabular}
\end{table}

\begin{figure*}[tbp]
\centerline{\includegraphics[scale=0.35]{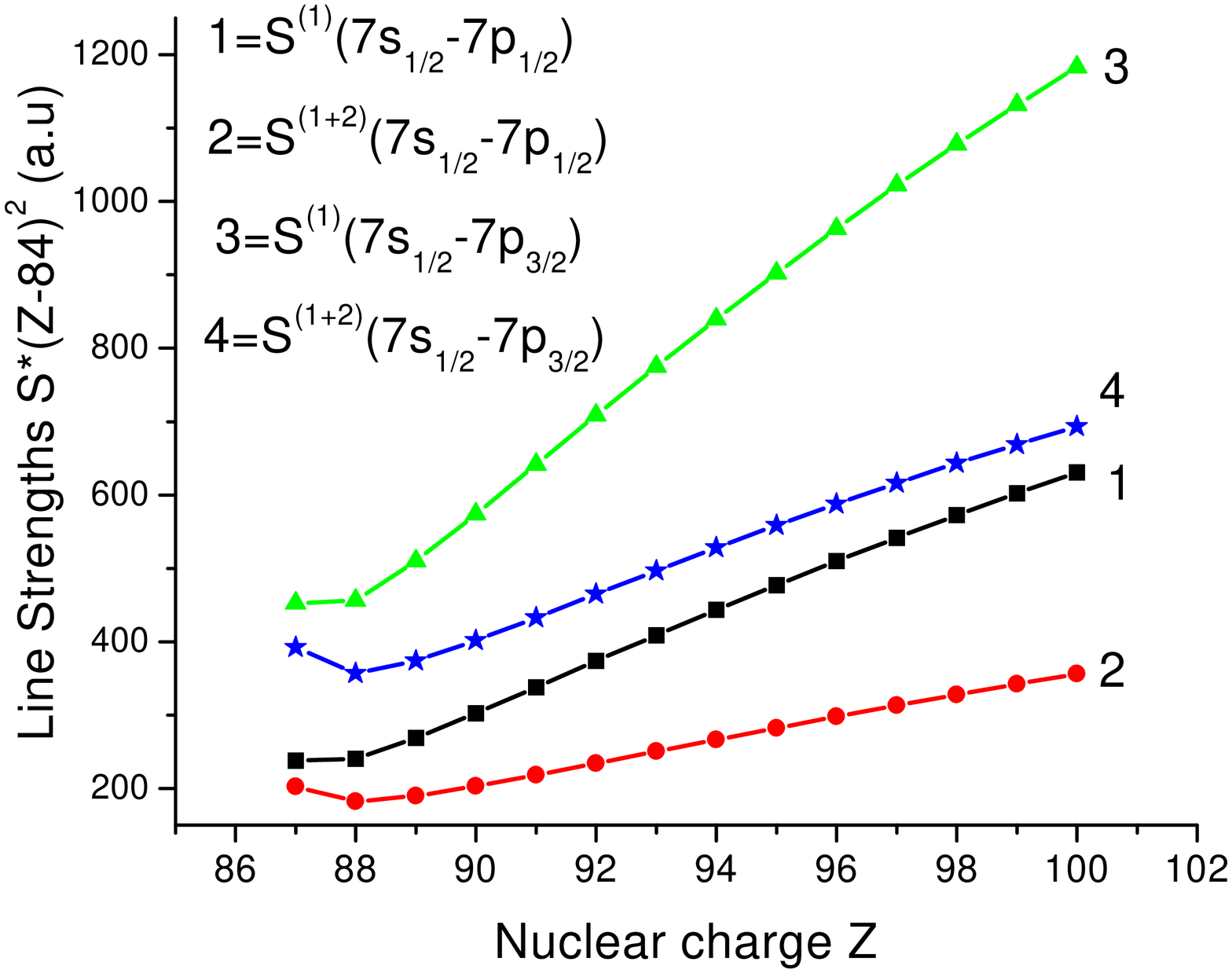}
           \includegraphics[scale=0.35]{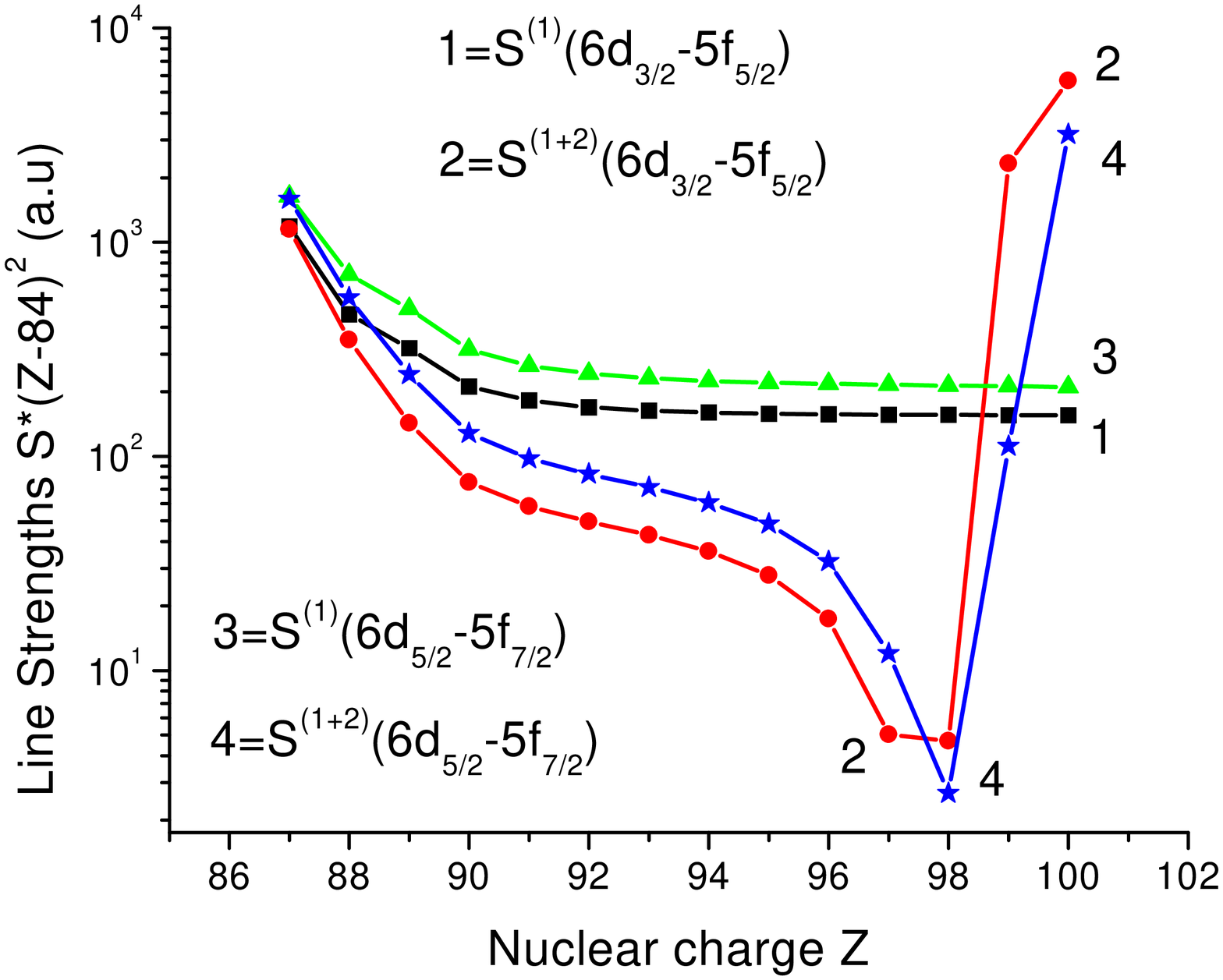}}
\centerline{\includegraphics[scale=0.35]{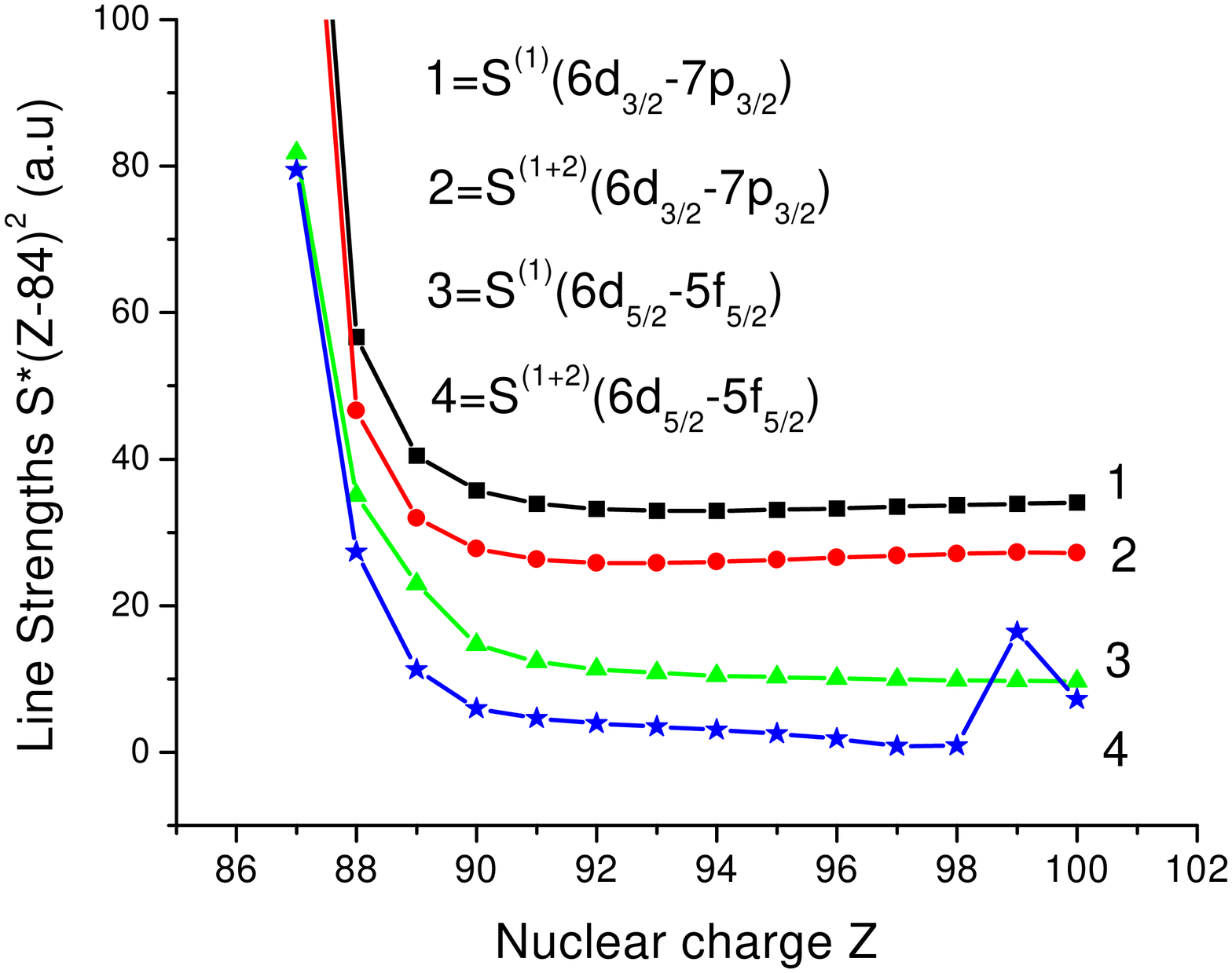}
           \includegraphics[scale=0.35]{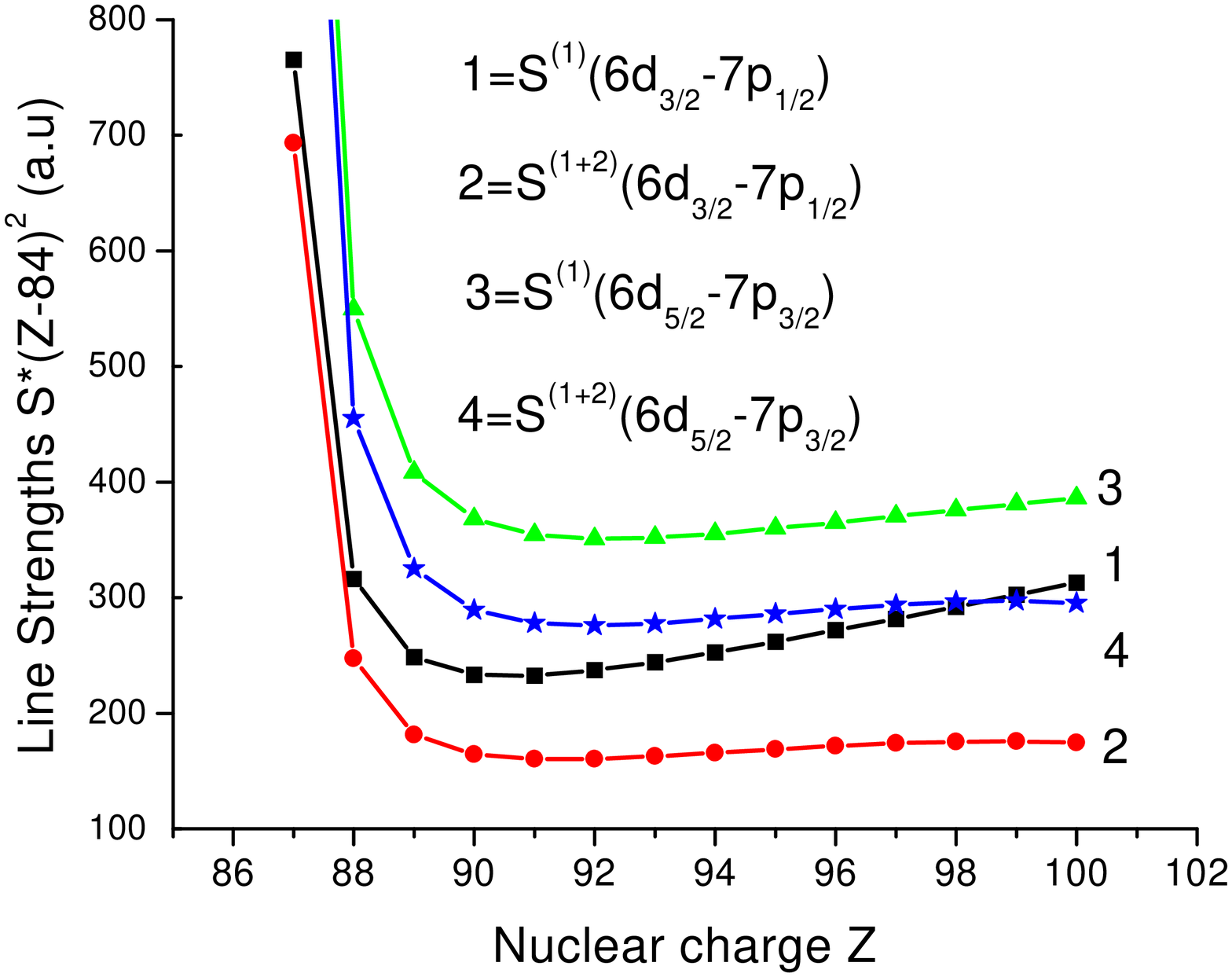}}
\caption{Line Strengths  ($S\times (Z-84)^2$ in a.u)
 as functions of $Z$  in Fr-like ions. }
 \label{fig-s1}
\end{figure*}

\section{Electric-dipole matrix elements, oscillator strengths, transition
rates, and lifetimes in Fr-like ions}

\subsection{Electric-dipole matrix elements}

The matrix element of a one-particle operator $Z$ is given by
\cite{blundell-li}
\begin{equation}\label{eq11}
Z_{wv}=\frac{\left\langle \Psi _{w}\right| Z\left| \Psi _{v}\right\rangle }{%
\sqrt{\left\langle \Psi _{v}|\Psi _{v}\right\rangle \left\langle
\Psi _{w}|\Psi _{w}\right\rangle }}\,,
\end{equation}
where $\Psi _{v}$ is the exact wave function for the many-body
``no-pair'' Hamiltonian $H$. In MBPT, we expand the many-electron
wave function $\Psi _{v}$  in powers of $V_{I}$ as
\begin{equation}\label{eq13}
\left| \Psi _{v}\right\rangle =\left| \Psi _{v}^{(0)}\right\rangle
+\left| \Psi _{v}^{(1)}\right\rangle +\left| \Psi
_{v}^{(2)}\right\rangle +\left| \Psi _{v}^{(3)}\right\rangle +
\cdots\,.
\end{equation}
 The denominator in Eq.~(\ref{eq11}) arises
from the normalization condition that contributes starting from
third order \cite{equation}. In the lowest order,
we find
\begin{equation}\label{eq14}
Z_{wv}^{(1)}=\left\langle \Psi _{w}^{(0)}\right| Z\left| \Psi
_{v}^{(0)}\right\rangle =z_{wv}\,,
\end{equation}
where $z_{wv}$ is the  corresponding one-particle matrix element
\cite{multipol}. Since $\Psi _{w}^{(0)}$  is a DF function
we designate $Z^{(1)}$ by $Z^{(\rm DF)}$  below.

The second-order Coulomb correction to the transition matrix
element in the DF case with $V^{N-1}$ potential is given by
\cite{dip3}
\begin{equation}\label{eq15}
Z_{wv}^{(2)}=\sum_{na}\frac{z_{an}(g_{wnva}-g_{wnav})}{\varepsilon
_{a}+\varepsilon _{v}-\varepsilon _{n}-\varepsilon _{w}}+\sum_{na}
\frac{(g_{wavn}-g_{wanv})z_{na}}{\varepsilon _{a}+\varepsilon
_{w}-\varepsilon _{n}-\varepsilon _{v}}\,.
\end{equation}
Second-order Breit corrections are obtained from
Eq.~(\ref{eq15})
 by changing  $g_{ijkl}$ to $b_{ijkl}$, where $b_{ijkl}$ is the matrix element
 of the Breit operator given in ~\cite{li-en}.

In the all-order SD calculation,  we substitute the all-order SD
wave function $\Psi _{v}^{\rm SD}$  into the matrix element
expression given by Eq.~(\ref{eq11})  and obtain the expression
 ~\cite{blundell-li}
\begin{equation}\label{eq18}
Z_{wv}^{(\rm SD)}=\frac{z_{wv}+Z^{(a)}+ \cdots +Z^{(t)}}{\sqrt{(1+N_{w})(1+N_{v})}%
}\,,
\end{equation}
where $z_{wv}$ is the lowest-order  (DF) matrix element given
by Eq.~(\ref{eq14}),  and the terms $Z^{(k)}$, $k=a \cdots t$
are linear or quadratic functions of the
excitation coefficients introduced in Eq.~(\ref{eq8}). The
normalization terms $N_{w}$ are quadratic functions of the
excitation coefficients. As a result,  certain sets of many-body
perturbation theory terms are summed to all orders.  In contrast to
the energy, all-order SD matrix elements contain the entire
third-order MBPT contribution.

The calculation of the transition matrix elements provide another
test of the quality of atomic-structure calculations and another
measure of the size of correlation corrections. Reduced
electric-dipole matrix elements between low-lying states of
Fr-like systems with $Z$ = 87--92
 calculated in various  approximations are presented in Table~\ref{tab-dip}.

 Our  calculations of the reduced matrix elements in the lowest,
second, and third orders    are carried out following the method
described above. The
   lowest order DF value is obtained from Eq.~(\ref{eq14}).
 The values  $Z^{({\rm DF}+2)}$ are obtained as the sum of
 the second-order correlation correction $Z^{(2)}$ given by
Eq.~(\ref{eq15})  and  the DF matrix elements $Z^{(\rm DF)}$.  The
second-order Breit corrections $B^{(2)}$ are rather small in
comparison with  the second-order Coulomb correction $Z^{(2)}$
(the ratio of $B^{(2)}$ to $Z^{(2)}$ is about 0.2\%--2\%).

 The third-order matrix elements $Z^{({\rm DF}+2+3)}$ include the DF values,
 the second-order $Z^{(2)}$ results,
and the third-order $Z^{(3)}$  correlation correction. $Z^{(3)}$
includes random-phase-approximation  terms (RPA) iterated  to all
orders ~\cite{dip3}.

 We find correlation corrections  $Z^{(2+3)}$ to be  very
large, 10-25\%, for many cases. All results given in
Table~\ref{tab-dip} are obtained using length form of the  matrix
elements. Length-form and velocity-form matrix elements differ
typically by 5--20\% for the  DF matrix elements and 2--5~\% for
the second-order matrix elements in these calculations.

Electric-dipole matrix elements evaluated in the all-order SD
approximation are given  in  columns labeled $Z^\text{(SD)}$
(Eq.~(\ref{eq18})) of Table~\ref{tab-dip}. The SD matrix elements
$Z^\text{(SD)}$ include $Z^{(3)}$ completely, along with important
fourth- and higher-order corrections. The fourth-order corrections
omitted from the SD matrix elements were discussed recently by
\citet{der-4}.  The $Z^\text{{(SD)}}$ values  are smaller than the
$Z^{({\rm DF}+2)}$ values and larger than the $Z^{({\rm DF}+2+3)}$
values for all  transitions given in Table~\ref{tab-dip}.

In Fig.~\ref{fig-s1}, we illustrate the $Z$-dependences of the
line strengths for the $7s_{1/2} - 7p_{j}$, $6d_{j} - 7p_{j'}$,
and  $6d_{j} - 5f_{j'}$ transitions. Two sets of line strengths
values  $S^{(1)}$ and $S^{(1+2)}$  are presented for each
transition. The values of $S^{(1)}$ and $S^{(1+2)}$ are obtained
as  $(Z^{(\rm DF)})^2$ and $(Z^{(\rm DF+2)})^2$, respectively. It
should be noted that the values are scaled by $(Z-84)^2$  to
provide better presentation of the line strengths.  The difference
between $S^{(1)}$ and $S^{(1+2)}$ curves increases with increasing
$Z$; for the $7s_{1/2} - 7p_{1/2}$ transition, the ratio of the
second-order ($S^{(1+2)}$ - $S^{(1)}$) and the first-order
($S^{(1)}$) contributions is equal to 15\% and 37\% for $Z$ =87
and 92, respectively.

\subsection{Form-independent third-order transition amplitudes}

We calculate electric-dipole reduced matrix elements using the
form-independent third-order perturbation theory developed by
Savukov and Johnson in Ref.~\cite{savukov01}. Previously, a good
precision of this method has been demonstrated for alkali-metal
atoms. In this method, form-dependent ``bare''  amplitudes are
 replaced with  form-independent
random-phase approximation (``dressed'') amplitudes to obtain
form-independent third-order amplitudes to some degree of
accuracy. As in the case of the third-order energy calculation, a
limited number of partial waves with $l_{\max }<7$ is included.
This restriction is not very important for considered here ions
because third-order correction is quite small, but it gives rise
to some loss of gauge invariance. The gauge independence serves as
a check that  no numerical problems occurred.

Length and velocity-form matrix elements from DF, second-order
(RPA), and third-order calculations are given in
Table~\ref{tab-LV} for the limited number of transitions in
Fr-like systems with $Z$ = 87--100.
  The $Z^{(\rm DF)}$ values differ in $L$ and $V$ forms by
2--15~\% for the $p-s$ transitions. The very large  $L-V$
difference (by a factor of 2--3) is observed in the $Z^{(\rm DF)}$
values  for $d-f$ transitions  as illustrated in
Table~\ref{tab-LV}. The second-order RPA contribution removes this
difference in $L-V$ values, and  the $L$ and $V$ columns with the
$Z^{(\rm DF +2)}$ headings are almost identical. There are,
however, small $L-V$ differences (0.002\%--0.2\%)  in the
third-order matrix elements. These remaining  small differences
can be explained by limitation in the number of partial waves
taken into account in the $l_{max}$ in the third-order matrix
element evaluations that we already discussed in the previous
section describing the energy calculations.

\subsection{Oscillator strengths, transition rates, and lifetimes}

We calculate  oscillator strengths and transition probabilities
for the  eight $7s-7p$, $7p-6d$, and $6d-5f$ electric-dipole
transitions in Ra~II, Ac~III, Th~IV, U~VI and for the  seven
$7s-7p$, $7p-7d$, and $7p-8s$ electric-dipole transitions in Fr~I.
Wavelengths $\lambda$ (\AA), weighted transition rates $gA$
(s$^{-1}$), and oscillator strengths $gf$ in Ac~III, Th~IV, U~VI
are given in Table~\ref{tab-com1} for Ac~III, Th~IV, U~VI ions and
in Table~\ref{tab-com2} for Ra~II and Fr~I.

 The SD data
 ($gA^{\mathrm{(SD)}}$ and $gf^{\mathrm{(SD)}}$) are compared with
  theoretical ($gA^{\mathrm{(RHF)}}$ and $gf^{\mathrm{(RHF)}}$)
  results from
   Ref.~\cite{osc-ra}.
  The  experimental energies were used
  to calculate the $gA^{\mathrm{(SD)}}$, $gf^{\mathrm{(SD)}}$,
$gA^{\mathrm{(RHF)}}$, and $gf^{\mathrm{(RHF)}}$. Therefore, we
really compare the values of the electric-dipole matrix elements.
The SD and RHF results for $s-p$ and  $p-d$ transitions disagree
by about 6--25~\% with  exception of  the $6d_{j} - 7p_{3/2}$
transitions where disagreement is   60~\%. The largest
disagreement (by a factor of 2--5) is observed  between the SD and
 RHF results for the  $f-d$ transitions. The correlation
 corrections  are especially large for these transitions as
 illustrated in Table~\ref{tab-dip}. Therefore, we expect that
 our values, which include correlation correction in rather
 complete way, will disagree with RHF calculations which appear  not
 to include any correlation effects for transitions which involve
 $5f$ states.
Our conclusion is confirmed by
 comparison of the  $gA^{\mathrm{(RHF)}}$ and $gf^{\mathrm{(RHF)}}$
 and our $gA^{\mathrm{(DF)}}$ and $gf^{\mathrm{(DF)}}$ results (see, also Ref~\cite{thiv}).
Our values for transitions rates
 and oscillator strengths are in reasonable agreement (10--20~\%)
with RHF data.

The SD data
 $gA^{\mathrm{(SD)}}$ and $gf^{\mathrm{(SD)}}$ for Fr~I and Ra~II
 given Table~\ref{tab-com2} are compared with
  theoretical data ($gA^{\mathrm{(RHF)}}$,  $gf^{\mathrm{(RHF)}}$)
  given in Refs.~\cite{biemont-fr,osc-ra}  and
  theoretical ($gA^{\mathrm{(MBPT)}}$ and  $gf^{\mathrm{(MBPT)}}$) values from
   Ref.~\cite{dzuba-01}.  The
  wavelengths $\lambda^\text{{(expt)}}$ given in Table~\ref{tab-com2}
are taken  from  NIST compilation \cite{web-nist} for Ra~II and
from Ref.~\cite{expt} ($E(7p_{1/2})$ = 12237.409~cm$^{-1}$,
$E(7p_{3/2})$ = 13923.998~cm$^{-1}$), Ref.~\cite{fr-8s}
($E(8s_{1/2})$ = 19732.523~cm$^{-1}$), and Ref.~\cite{fr-7d}
($E(7d_{3/2})$ = 24244.831~cm$^{-1}$ and $E(7d_{5/2})$ =
24333.298~cm$^{-1}$) for Fr~I. Comparison of the $gf$ and $gA$
results obtained by three different approximations shows better
agreement between our SD and MBPT \cite{dzuba-01} results than
between our and  HFR results obtained Bi\'{e}mont {\it et al.\/}
in Refs.~\cite{osc-ra,biemont-fr}, owing to more complete
treatment of the correlation  in our calculation and in
Ref.~\cite{dzuba-01}.

 Our SD lifetime results are compared in Table~\ref{tab-life}
with experimental
 measurements presented in Refs.~\cite{fr-98,fr-00,fr-05} for the
 $7p_{j}$, $7d_j$, $8s_{1/2}$ levels in neutral francium. We find
  that our SD lifetimes are in excellent
 agreement with precise measurements provided in
 Refs.~\cite{fr-98,fr-00,fr-05}.

\begin{table*}
\caption{\label{tab-mult} Reduced matrix elements of the
electric-quadrupole (E2)
 and magnetic-multipole (M1 and M3) operators in first, second,
third, and all orders of perturbation theory in  Fr-like ions.}
\begin{ruledtabular}
\begin{tabular}{lllrrrr}
\multicolumn{1}{c}{}& \multicolumn{2}{c}{Transition}&
\multicolumn{1}{c}{$Z^{({\rm DF})}$ }&
\multicolumn{1}{c}{$Z^{({\rm DF}+2)}$ }&
\multicolumn{1}{c}{$Z^{({\rm DF}+2+3)}$ }&
\multicolumn{1}{c}{$Z^\text{{(SD)}}$ }\\
\hline
   \multicolumn{7}{c}{Ra~II, $Z$ = 88}\\
M3 &$6d_{5/2}$ &$   7s_{1/2}$&   128.8000&    131.1100&    128.7695&    120.0000\\
E2 &$6d_{3/2}$ &$   7s_{1/2}$&    17.2630&     17.0350&     13.7518&     14.5885\\
E2 &$6d_{5/2}$ &$   7s_{1/2}$&    21.7710&     21.5580&     17.8089&     18.6906\\
M1 &$6d_{3/2}$ &$   7s_{1/2}$&     0.0000&      0.0008&      0.0189&      0.0014\\
   \multicolumn{7}{c}{Ac~III, $Z$ = 89}\\
M3 &$6d_{5/2}$ &$   7s_{1/2}$&    80.7820&     82.9520&     81.2050&     76.2511\\
E2 &$6d_{3/2}$ &$   7s_{1/2}$&    10.6820&     10.4410&      9.2237&      9.5149\\
E2 &$6d_{5/2}$ &$   7s_{1/2}$&    13.6550&     13.4270&     11.9668&     12.2808\\
M1 &$6d_{3/2}$ &$   7s_{1/2}$&     0.0000&      0.0015&      0.0203&      0.0013\\
   \multicolumn{7}{c}{Th~IV, $Z$ = 90}\\
M1  &$5f_{5/2}$ &$   5f_{7/2}$&     1.8506&      1.8525&      1.8390&      1.8514\\
E2  &$5f_{5/2}$ &$   5f_{7/2}$&     1.5669&      1.2339&      0.9724&      1.0834\\
   \multicolumn{7}{c}{Pa~V, $Z$ = 91}\\
M1  &$5f_{5/2}$ &$   5f_{7/2}$&     1.8505&      1.8520&      1.8421&      1.8513\\
E2  &$5f_{5/2}$ &$   5f_{7/2}$&     1.1348&      0.8203&      0.7033&      0.7613\\
   \multicolumn{7}{c}{U~VI, $Z$ = 92}\\
M1  &$5f_{5/2}$ &$   5f_{7/2}$&     1.8505&      1.8518&      1.8443&      1.8512\\
E2  &$5f_{5/2}$ &$   5f_{7/2}$&     0.9153&      0.6205&      0.5413&      0.5864\\
\end{tabular}
\end{ruledtabular}
\end{table*}

\begin{figure*}[tbp]
\centerline{\includegraphics[scale=0.35]{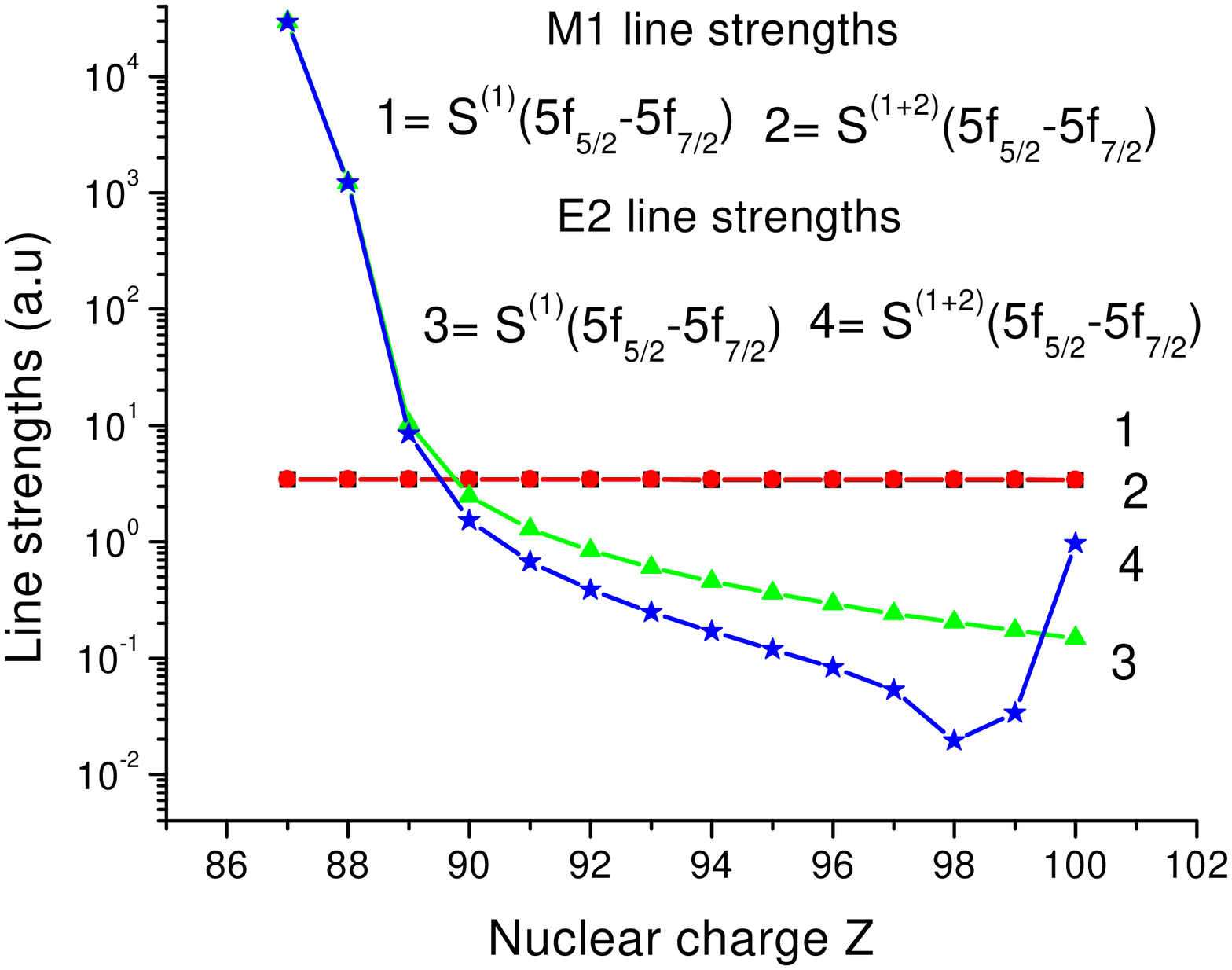}
           \includegraphics[scale=0.35]{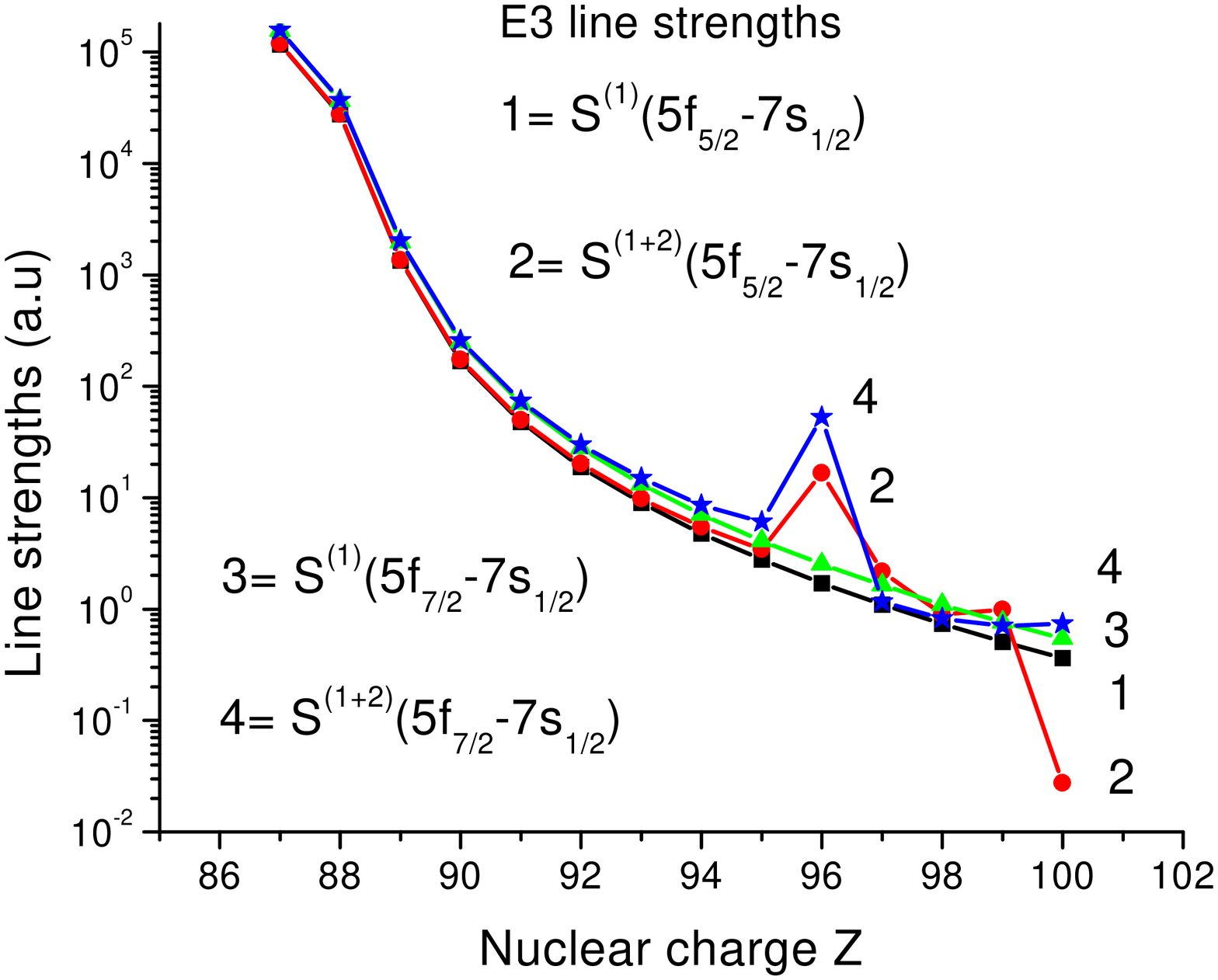}}
\centerline{\includegraphics[scale=0.35]{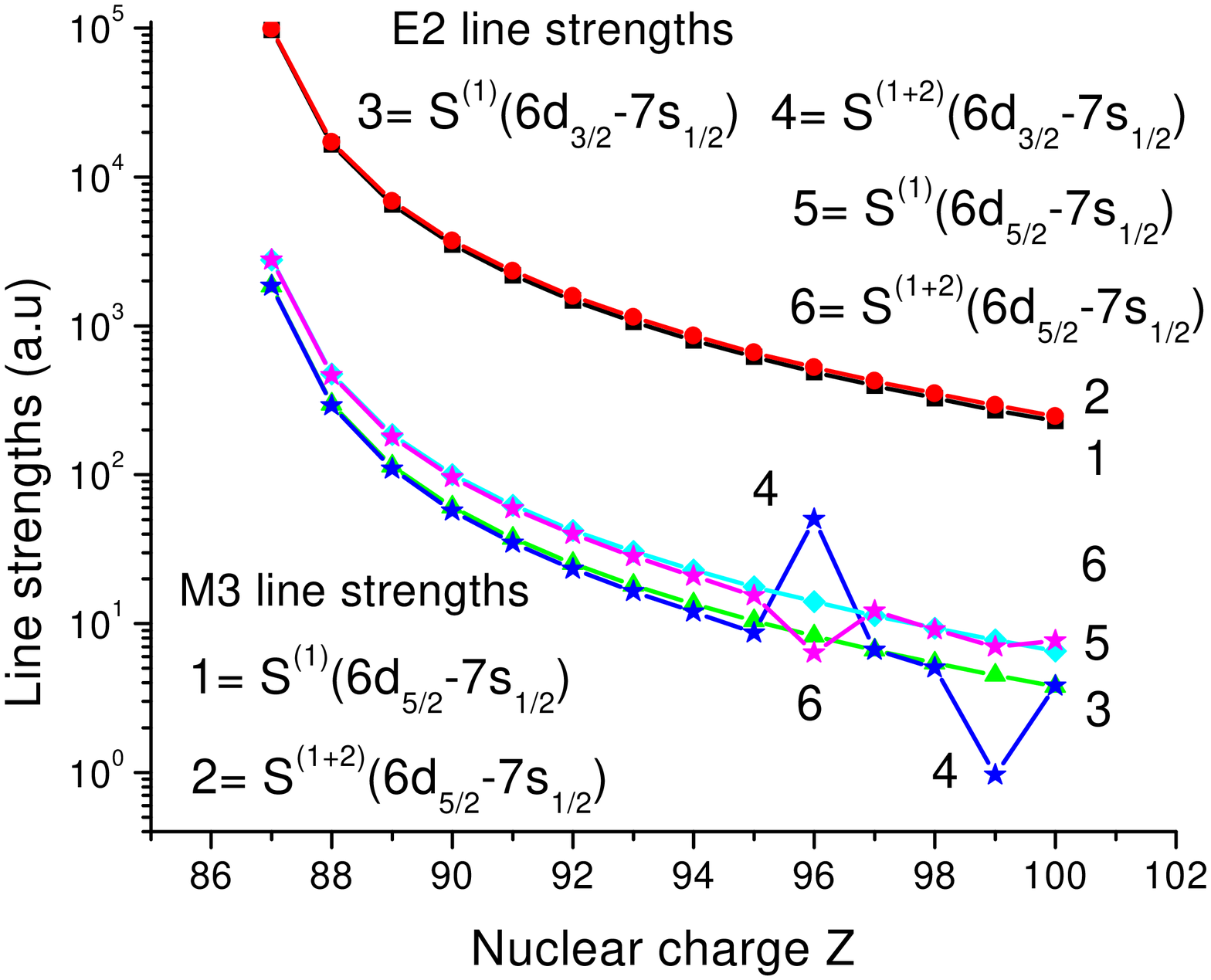}
            \includegraphics[scale=0.35]{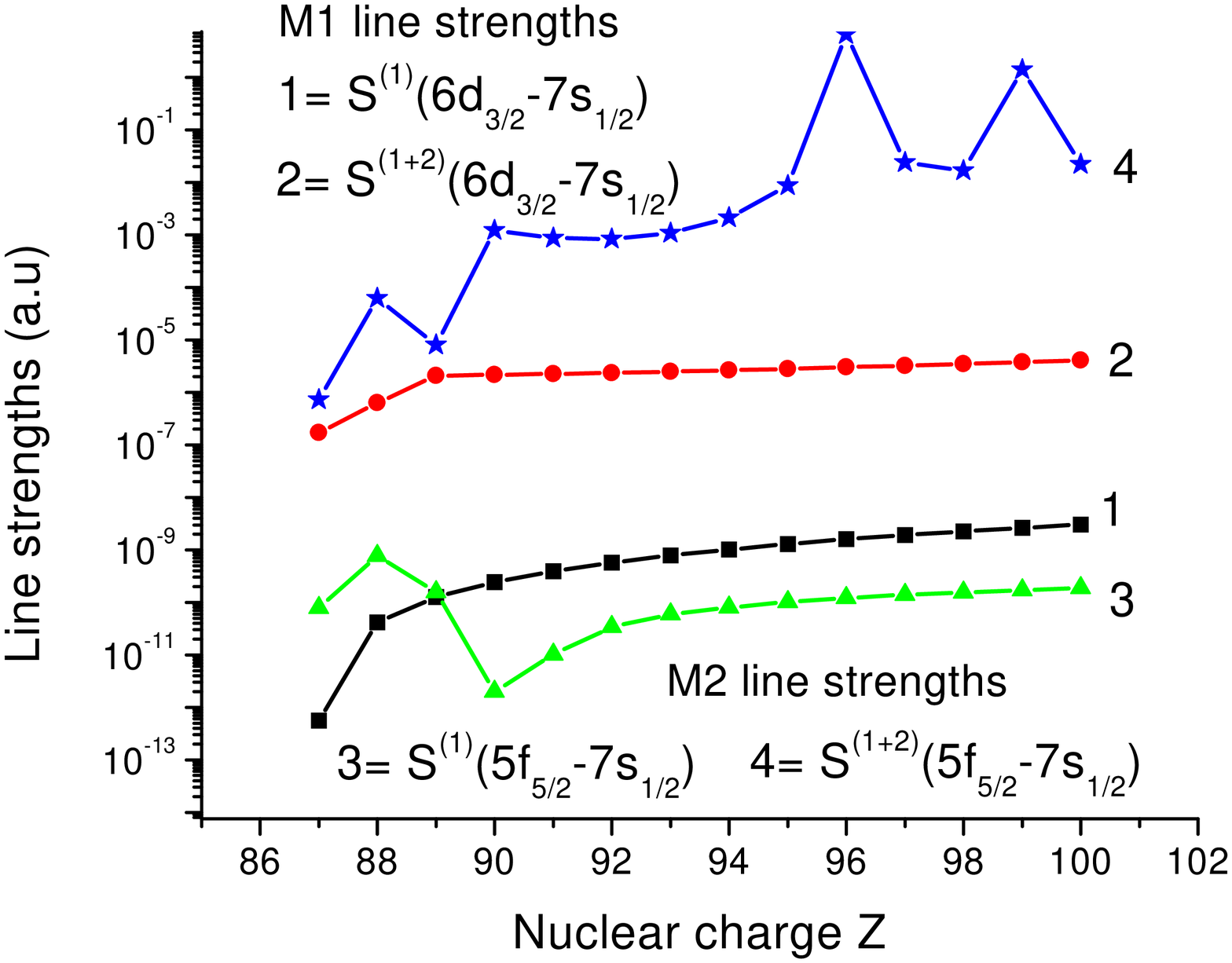}}
\caption{Multipole line strengths  ($S$ in a.u)
 as functions of $Z$  in Fr-like ions. }
 \label{fig-mult}
\end{figure*}

\begin{table}
\caption{\label{tab-ar-mult} Wavelengths $\lambda$ (\AA) and
transition rates $A^{(SD)}$
 (s$^{-1}$)  of the electric-quadrupole  (E2)
 and magnetic-multipole (M1 and M3)
  transitions in Ra~II, Ac~III, Th~IV, and U~VI  calculated in
the SD approximation. Numbers in brackets represent powers of 10.}
\begin{ruledtabular}
\begin{tabular}{lllrlll}
\multicolumn{1}{c}{} & \multicolumn{2}{c}{Transition} &
\multicolumn{1}{c}{$\lambda$} & \multicolumn{1}{c}{$A^{\rm
(SD)}$}& \multicolumn{1}{c}{$\lambda$} &
\multicolumn{1}{c}{$A^{\rm (SD)}$} \\
\hline \multicolumn{3}{c}{} &
   \multicolumn{2}{c}{Ra~II, $Z$ = 88}&
   \multicolumn{2}{c}{Ac~III,$Z$ = 89}\\
E2 &$7s_{1/2}$ &$6d_{3/2}$&       8275.145    &1.536[+00]&       124844.      &1.053[-06]\\
E2 &$7s_{1/2}$ &$6d_{5/2}$&       7276.373    &3.197[+00]&       23787.4      &3.696[-03]\\
M3 &$7s_{1/2}$ &$6d_{5/2}$&       7276.373    &6.988[-07]&       23787.4      &7.070[-11]\\
M1 &$7s_{1/2}$ &$6d_{3/2}$&        8275.145   &2.281[-05]&        124844.     &4.354[-13]\\
\hline \multicolumn{3}{c}{} &
   \multicolumn{2}{c}{Th~IV, $Z$ = 90}&
   \multicolumn{2}{c}{U~VI, $Z$ = 90}\\
M1 &$5f_{5/2}$ &$5f_{7/2}$&     23119.6      &9.352[-01]&     13143.0      &5.089[+00]\\
E2 &$5f_{5/2}$ &$5f_{7/2}$&     23119.6      &2.487[-05]&     13143.0      &1.227[-04]\\
\end{tabular}
\end{ruledtabular}
\end{table}

\section{Multipole matrix elements, transition rates, and
lifetimes in Fr-like ions}

Reduced matrix elements of the electric-quadrupole (E2) and
magnetic-multipole (M1, M2, and M3) operators in lowest, second,
third, and all orders of perturbation theory are given in
Table~\ref{tab-mult} for Fr-like ions with $Z$ = 88--92. Detailed
descriptions of the calculations  of the  multipole matrix
elements
 in lowest and second orders of
perturbation theory were given  in
Refs.~\cite{multipol,ca,cjp-04}. Third-order and all-order
calculations are carried out using  the same method as the
calculations of E1 matrix elements. In Table~\ref{tab-mult}, we
present  E2, M1, and M3 matrix elements in the $Z^{(\rm DF)}$,
$Z^{(\rm DF+2)}$, $Z^{(\rm DF+2+3)}$, and $Z^\text{(SD)}$
approximations for the  $6d_{j} - 7s_{1/2}$ transitions in Ra~II
and  Ac~III and the M1, E2 $5f_{5/2} - 5f_{7/2}$ transition in
Th~IV, Pa~V, and U~VI. We already mentioned that the ground state
in Ra~II and Ac~III is the $7s_{1/2}$ state with the $6d_{j}$
being the next lowest  states; however, the ground state in Th~IV,
Pa~V, and U~VI  is the $5f_{3/2}$ state with the $5f_{5/2}$ being
the next lowest state.
 The
second-order contribution is about 1--3\% for all transitions
involving the $7s_{1/2}$ states. For the $5f_{5/2} - 5f_{7/2}$
transition, the second-order contribution (Coulomb and Breit)  is
very small (0.1\%) for the M1 transition and is rather large
(20\%) for the E2 transition.

In Fig.~\ref{fig-mult}, we illustrate the $Z$-dependences of the
line strengths for the $5f_{5/2} - 5f_{7/2}$, $6d{j} - 7s_{1/2}$,
and  $5f_{j} - 7ps_{1/2}$ transitions. Two sets of line strengths
 calculated in first- and second-order approximations
 are presented for each transition.
The values of $S^{(1)}$ and $S^{(1+2)}$ are obtained as  $(Z^{(\rm
DF)})^2$ and $(Z^{(\rm DF+2)})^2$, respectively.  The difference
between $S^{(1)}$ and $S^{(1+2)}$ curves increases with increasing
$Z$. For the $5f_{5/2} - 5f_{7/2}$ transition, the ratio of the
second-order ($S^{(1+2)}$ - $S^{(1)}$) and the first-order
($S^{(1)}$) contributions is equal to 18\% and 67\% for $Z$ =89
and 95, respectively.

The  strong irregularities occur in the curves describing the
second-order contributions (see, for example, $S^{(1+2)}(5f_{j} -
7s_{1/2}))$ for $Z$ = 96 and  $S^{(1+2)}(6d_{3/2} - 7s_{1/2}))$
for $Z$ = 95, 99). Those sharp features are explained  by
accidentally small energy denominators in MBPT expressions for
correlation corrections as discussed above.

 Wavelengths and transition rates $A^{(SD)}$
 for  the electric quadrupole (E2) and magnetic-multipole (M1 and M3)
  transitions  in Ra~II, Ac~III, Th~IV, and U~VI  calculated in
the SD approximation are
  presented in Table~\ref{tab-ar-mult}.
 The largest contribution to the lifetime of the $6d_{j}$
  state in Ra~II and Ac~III ions
  comes from the E2 transition, but
   the largest contribution to the lifetime of the $5f_{7/2}$
  state in Th~IV ~\cite{thiv}  and U~VI ions comes from the M1 transition.
  Our SD result for the M1 matrix element
   in Th~IV ~\cite{thiv} ion  agrees to 0.5\%  with HFR results obtained by
 Bi\'{e}mont {\it et al.\/} in Ref.~\cite{osc-ra} since the correlation
 is small  for M1 transition.

Finally, we find that the lifetimes of the $6d_{3/2}$ state is
equal to 0.651~s in Ra~II and 9.50$\times$10$^5$~s in Ac~III;  the
lifetime of the $6d_{3/2}$ state is equal to 0.313~s in Ra~II and
271~s in Ac~III.
  The lifetime of the $5f_{7/2}$ state is equal to
1.07~s in Th~IV and 0.196~s in U~VI.

\section{ Ground state static polarizabilities for Fr-like ions}

The static polarizability of Fr-like ions can be calculated as the
sum of the polarizability of the ionic core $\alpha_c$, a counter
term
 $\alpha_{vc}$ compensating for excitations from the core to the
 valence shell which violate the Pauli principle, and a valence
 electron contribution $\alpha_v$:
\begin{equation}\label{eq-p1}
\alpha =\alpha _{\mathrm{c}}+\alpha _{\mathrm{v}}+\alpha _{%
\mathrm{vc}}.
\end{equation}
These contributions are given by formulas listed, for example, in
Refs.~\cite{st-70,mar-pol-04}.
 We calculate
$\alpha _{\mathrm{c}}$ in the relativistic RPA approximation (see
Ref.~\cite{adndt-83}).

  The $7s_{1/2}$ is the ground state in the
cases of Fr~I, Ra~II, and Ac~III, and the corresponding
polarizability terms are given by
\begin{eqnarray}\label{eq-p3}
\alpha _{\mathrm{v}}
&=&\sum_{n=7}^{N}[I_{v}(np_{1/2})+I_{v}(np_{3/2})],\quad
\\\nonumber \alpha _{\mathrm{vc}}
&=&\sum_{n=2}^{6}[I_{v}(np_{1/2})+I_{v}(np_{3/2})]\,,
\end{eqnarray}
where
\begin{equation}\label{eq-p4}
I_{v}(nlj)=\frac{2}{3(2j+1)}\frac{(Z_{v,nlj})^{2}}{E_{nlj}-E_{v}}\,,
\end{equation}
and N is the size of B-spline basis set (N = 50 in this
calculation).
 The calculation of the $\alpha _{v}$ is
divided into two parts:
\begin{eqnarray}  \label{main}
\alpha _{v}^{\mathrm{main}}
&=&\sum_{n=7}^{k}[I_{v}(np_{1/2})+I_{v}(np_{3/2})];\   \nonumber
\\
\alpha _{v}^{\mathrm{tail}}
&=&\sum_{n=k+1}^{N}[I_{v}(np_{1/2})+I_{v}(np_{3/2})].
\end{eqnarray}
Here, $k$ is equal to 10, 9, and 8 for Fr~I, Ra~II, and Ac~III.
The values of $\alpha _{v}^{\mathrm{main}}$ are calculated using
SD values of dipole matrix elements $Z_{v,nlj}$ and experimental
energies where they are available. We use  SD energies when we did
not find experimental data. The $\alpha _{v}^{\mathrm{tail}}$. and
$\alpha _{\mathrm{c}}$ contributions are small and are calculated
in DF approximation.

Our numerical results are given in Table~\ref{tab-pol-7s}. The sum
over $n$  in the main polarizability term given by
Eq.~(\ref{main}) converges faster for Ac~III than for Fr~I.
 The ratios of the second and first terms in sum over $n$
 are equal to 1\% for Fr~I  and only 0.1\% for Ac~III. Therefore,
 fast
convergence allows us to limit the number of $n$ in
Eq.~(\ref{main})  to $n$ = 8 in Ac~III, since we have no
experimental energy values for high $n$ for this ion.

Our SD results for $\alpha_{7s}^{\rm SD}$   given in last line of
Table~\ref{tab-pol-7s} are in good agreement with recommended
value (317.7$\pm$2.4) for Fr given by Derevianko {\it et al.\/} in
Ref.~\cite{fr-pol} and recommended value (104.0) for Ra~II given
recently by Lim and P. Schwerdtfeger in Ref.~\cite{ra-pol}. We did
not find  any  data for the $\alpha_{7s}^{\rm SD}$ in Ac~III. Our
ionic core polarizabilities $\alpha _c$ given in
Table~\ref{tab-pol-7s} are in an excellent agreement with
recommended values (20.4 in Fr~I and 13.7 in Ra~II) presented in
Refs.~\cite{fr-pol,ra-pol}.

The valence polarizability for  the $5f_{5/2}$ state, which is the
ground state in the case of Fr-like ions with $Z \geq$ 90, is
given by ~\cite{thiv}
\begin{equation}\label{eq-p6}
\alpha _{{\rm
v}}=\sum_{n=n_0}^{N}[I_{v}(nd_{3/2})+I_{v}(nd_{5/2})+I_{v}(ng_{7/2})].
\end{equation}
Here, $n_0$ equal to 6 for the $nd$ states and 5 for the $ng$
states.

 We use the same designations as  we use  for the $7s_{1/2}$ polarizability
 given by Eqs.~(\ref{eq-p1})~-~(\ref{main}). Our results are listed in Table~\ref{tab-pol-th}.
The third-order (DF+2+3) and SD dipole matrix elements (a.u.) are
calculated with the 50
 splines and  cavity radius $R$ = 45 for Th~IV.
  We use use experimental energies \cite{expt}  to calculate
 $\alpha _{v}^{\mathrm{main}}$ in Table~\ref{tab-pol-th}. Both,
 third-order and all-order results for
  dipole polarizability of Th~IV  in the ground
  state $5f_{5/2}$  are presented in Table~\ref{tab-pol-th}
  (see also Ref.~\cite{thiv}). We  see from
this table that the largest contribution to the $\alpha
_{v}^{\mathrm{main}}$ term comes from the $6d$ states ($\alpha
_{v}^{\mathrm{main}}(6d)$).  The core contribution ($\alpha _c$ =
7.750~a$_0$$^3$) is larger than the main term, limiting the
accuracy of our calculations. The $\alpha _{v}^{\mathrm{tail}}$
and $\alpha _{vc}$ terms calculated in the DF approximation
contribute only 5\% to the final results.  No experimental data
are available for Th~IV polarizability.

The $5f_{5/2}$ ground state  polarizabilities  for Fr-like ions
with $Z$ = 91--100 are given in Table~\ref{tab-pol-5f}. Results
are obtained in DF and second-order MBPT approximations.   The
contribution of the $5g_{7/2}$ state into the $\alpha _{5f_{5/2}}$
polarizability increases with increasing $Z$; 1\% for ion with $Z$
= 90 (Th~IV), 19\% for ion with $Z$ = 92 (U~VI), and 30\% for ion
with $Z$ = 95 (Am~IX). The main contributions are smaller than the
core contributions. The calculations are conducted only in
second-order approximation for these ions, owing to problems with
accidentally very small denominators in the corresponding
calculations of the properties with involving  $5g$ state. We
mentioned previously  that the core values $\alpha
_{\mathrm{c}}(c)$ are calculated in the relativistic RPA
approximation following by method described
 by Johnson {\it et al.\/} in
Ref.~\cite{adndt-83}.  Our $\alpha _{\mathrm{c}}(c)$ results for
Ra~II -- U~VI disagree with results presented by Bi\'{e}mont {\it
et al.\/} in Ref.~\cite{osc-ra}, but agree with Ra~II result
given  in Ref.~\cite{ra-pol}.

\section{Conclusion}
In summary,  a systematic relativistic  MBPT study of the atomic
properties
 of  the  $7s_{1/2}$, $7p_j$, $6d_j$, and $5f_j$ states in
 Fr-like ions with nuclear charges $Z = 87 - 100$ is presented.
 The energy values
are in good agreement with available  experimental energy data and
provide a theoretical reference database for the line
identification. A systematic
 all-order SD study of the reduced matrix elements and transition rates for
eight $7s-7p$, $7p-6d$, and $6d-5f$ electric-dipole transitions is
conducted. Multipole matrix elements ($7s_{1/2}\ - 6d_j$,
$7s_{1/2}\ - 5f_j$, and $5f_{5/2}\ - 5f_{7/2}$)  are evaluated to
obtain the lifetime data for the $6d_{3/2}$  and the $5f_{7/2}$
excited state. The scalar  polarizabilities for the $7s_{1/2}$
ground state in Fr~I, Ra~II, and Ac~III and $5f_{5/2}$ ground
state in Th~IV
 are calculated using a
relativistic third-order and all-order methods. The scalar
polarizabilities  for Fr-like ions with nuclear charge $Z$ =
90--100 in the $5f_{5/2}$ ground state
 are calculated using a
relativistic second-order  MBPT.
 These calculations provide a
theoretical benchmark for comparison with experiment and theory

\begin{acknowledgments}
The work of W.R.J. and U.I.S. was supported in part by National
Science Foundation Grant No.\ PHY-04-56828. The work of M.S.S. was
supported in part by National Science Foundation Grant  No.\
PHY-04-57078.
\end{acknowledgments}

\begin{table}
\caption{\label{tab-pol-7s} Contribution to the $7s_{1/2}$ static
polarizabilities (a.u.) of Fr~I, Ra~II, and Ac~III. }
\begin{ruledtabular}
\begin{tabular}{lrrr}
\multicolumn{1}{c}{$v=7s_{1/2}$}& \multicolumn{1}{c}{Fr~I}&
\multicolumn{1}{c}{Ra~II}&
\multicolumn{1}{c}{Ac~III}\\
\hline
$\alpha ^{{\rm main}}_{v}(7p)$   & 289.24    &       93.08    &     46.74\\[0.3pc]
$\alpha ^{{\rm main}}_{v}(8p)$   &   3.02    &        0.23    &      0.06\\[0.3pc]
$\alpha ^{{\rm main}}_{v}(9p)$   &   0.54    &        0.03    &          \\[0.3pc]
$\alpha ^{{\rm main}}_{v}(10p)$  &   0.19    &                &         \\ [0.3pc]
$\alpha ^{{\rm main}}_{v}$       & 292.99    &       93.34    &     46.80\\[0.4pc]
$\alpha ^{{\rm tail}}_{v}$       &   1.20    &        0.11    &      0.04\\[0.3pc]
$\alpha _c$                      &  20.40    &       13.79    &     11.42\\[0.3pc]
$\alpha _{vc}$                   &  -0.95    &       -0.74    &     -0.54\\[0.4pc]
$\alpha ^{{\rm SD}}_{v}$         &  313.7    &       106.5    &     57.71\\
\end{tabular}
\end{ruledtabular}
\end{table}

\begin{table}
\caption{\label{tab-pol-th} Contribution to the $5f_{5/2}$ static
polarizabilities (a.u.) of  Th~IV calculated in the third-order -
$(a)$ and all-order - $(b)$ approximations. }
\begin{ruledtabular}
\begin{tabular}{lrr}
\multicolumn{1}{c}{$v=5f_{5/2}$}&
\multicolumn{1}{c}{$\alpha ^{(a)}$}&
\multicolumn{1}{c}{$\alpha ^{(b)}$}\\
\hline
$\alpha ^{{\rm main}}_{v}(6d)$         & 4.961  & 6.491  \\[0.3pc]
$\alpha ^{{\rm main}}_{v}(7d)$         & 0.009  & 0.015  \\[0.3pc]
$\alpha ^{{\rm main}}_{v}(8d)$         & 0.005  & 0.007  \\[0.3pc]
$\alpha ^{{\rm main}}_{v}(5g_{7/2})$   & 0.063  & 0.078  \\[0.3pc]
$\alpha ^{{\rm main}}_{v}(6g_{7/2})$   & 0.023  & 0.023  \\[0.3pc]
\hline
$\alpha ^{{\rm main}}_{v}$             & 5.062  & 6.612  \\[0.3pc]
$\alpha ^{{\rm tail}}_{v}$             & 0.762  & 0.762  \\[0.3pc]
$\alpha _c$                            & 7.750  & 7.750  \\[0.3pc]
$\alpha _{vc}$                         & -0.050 &-0.050  \\[0.3pc]
$\alpha _{v}$                          & 13.52  & 15.07  \\[0.3pc]
\end{tabular}
\end{ruledtabular}
\end{table}

\begin{table*}
\caption{\label{tab-pol-5f} Contribution to the $5f_{5/2}$ static
polarizabilities (a.u.) in Fr-like ions with $Z$ =91-100
calculated in the DF - $(a)$, second-order - $(b)$, and RPA   -
$(c)$ approximations. }
\begin{ruledtabular}
\begin{tabular}{lrrrrrrrrrr}
\multicolumn{1}{c}{}&
\multicolumn{1}{c}{$Z$ = 91}&
\multicolumn{1}{c}{$Z$ = 92}&
\multicolumn{1}{c}{$Z$ = 93}&
\multicolumn{1}{c}{$Z$ = 94}&
\multicolumn{1}{c}{$Z$ = 95}&
\multicolumn{1}{c}{$Z$ = 96}&
\multicolumn{1}{c}{$Z$ = 97}&
\multicolumn{1}{c}{$Z$ = 98}&
\multicolumn{1}{c}{$Z$ = 99}&
\multicolumn{1}{c}{$Z$ = 100}\\
\hline
$\alpha ^{{\rm main}}_{v}(a)$&  3.072 & 1.035&  0.548 &  0.356&  0.258& 0.199& 0.158& 0.129& 0.106& 0.089\\[0.3pc]
$\alpha ^{{\rm main}}_{v}(b)$&  2.050 & 0.415&  5.935 &  0.187&  0.128& 0.083& 0.076&  0.237&  0.120&  0.089\\[0.3pc]
$\alpha ^{{\rm tail}}_{v}(a)$&  0.367 & 0.206&  0.119 &  0.067&  0.036& 0.018& 0.009&  0.005&  0.003&  0.010\\[0.3pc]
$\alpha _{vc}(a)$            & -0.048 &-0.044& -0.040 & -0.037& -0.033&-0.030&-0.028& -0.025& -0.023& -0.021\\[0.3pc]
$\alpha _c(a)$                &  7.236 & 5.977&  5.023 &  4.279&  3.687& 3.207& 2.813&  2.484&  2.207&  1.971\\[0.3pc]
$\alpha _c(c)$                &  6.166 & 5.029&  4.182 &  3.534&  3.024& 2.615& 2.281&  2.006&  1.776&  1.582\\[0.3pc]
$\alpha ^{{\rm (DF)}}_{v}$      &  10.627& 7.174&  5.649 &  4.665&  3.948& 3.394& 2.952&  2.593&  2.293&  2.049\\[0.3pc]
$\alpha ^{{\rm (DF+2)}}_{v}$    &  8.535 & 5.606&  10.197&  3.751&  3.155& 2.686& 2.338&  2.223&  1.876&  1.660\\[0.3pc]
\end{tabular}
\end{ruledtabular}
\end{table*}

\end{document}